**Optical quantum cascade laser based on a MIM structure**

*M.V. Davidovich*[*]

Saratov National Research University named after N.G. Chernyshevsky,

410012 Saratov, Russia

[*]davidovicvhmv@info.sgu.ru

**Abstract.** A quantum cascade laser based on a metal-insulator-metal structure is proposed, the heterostructure of which consists of diamond and metal films. The profile of the quantum potential in a quasi-periodic MIM heterostructure and the structure of metastable levels in quantum wells are investigated. A number of CCL configurations have been proposed. The possibility of generating radiation from the UV to THz range is theoretically shown.

***Keywords***: quantum cascade laser, MIM structures, resonant tunneling diode, THz generator

## 1. INTRODUCTION

Quantum cascade lasers (QQLs) are phenomenal devices based on quantum wells (QWs) that provide generation in the IR to THz ranges [1–6]. Such devices have from several QWs to several dozen QWs. Typically, QQLs use semiconductor structures such as AlInGaAs|InP, AlInAs|GaInAs, AlxGa1-xAs|GaAs [1], and a number of others. A major advantage of QQL is that the oscillation frequencies are determined not by the bandgap width, but by the configuration of the heterostructure [1–6] and the quantum potential (QP) profile. This opens up the possibility of obtaining oscillation in a wide frequency range, up to the UV range. The resonant tunneling diodes (RTDs) discovered shortly before QQL, which were built on configurations with one and two QWs [7–23], have very small negative differential resistances in magnitude, significantly smaller than the characteristic impedances of transmission lines, which makes it difficult to obtain oscillation or significant amplification when they are incorporated into transmission lines and resonators. The QQL heterostructure has many periods and QWs, operates at sufficiently high voltages, and, in addition to oscillation, can also provide a large negative differential resistance in magnitude. However, the advantage of QQL lies in its direct coherent generation. RTDs, like QQLs, which are among the active devices in the THz range, are mainly modeled as quantum heterostructures based on the solution of the Schrödinger equation or density functional theory, without taking into account the specific designs of their integration into transmission lines, waveguides, and resonators. These are very short nanoscale structures, whereas QQLs can be implemented on extended structures. The current–voltage characteristics (IVCs) of RTDs obtained through this modeling are mainly used to analyze RTDs as lumped elements for switching high-frequency circuits [1,15,20]. For the purpose of generating THz radiation, RTDs are mainly used as components in active antennas [1]. The antenna housing serves as a resonator. Generation occurs when the loss resistance in the antenna is less than the magnitude of the effective inserted active (negative) resistance of the RTD. Generation of higher harmonics is also possible [22]. This type of generation can be called classical, whereas QQL operates as a coherent quantum device [1]. Modeling of QQL and RTD in the literature is carried out using a number of approximations; in particular, a rectangular periodic QP linearly biased by an anode voltage is usually used. This frequency profile is called a Stark ladder [7]. Often, the QP of narrow barriers is described using delta functions [7,23].

In this paper, we propose a QQL based on a metal–insulator–metal (MIM) structure. The difference is that metals have a high Fermi energy (FE), ranging from several to 14.6 eV (beryllium) and a degenerate electron gas. Their band structure is quite simple, and the effective masses (EM) are close to the free electron mass me, while insulators (good dielectrics) also have a simple band structure: an empty band gap and a conduction band (CZ). The heights of the potential barriers in MIM QQL can be easily adjusted by selecting the dielectric constant (DC) of the insulators (barriers). This simplifies the production of QP and modeling. Traditional QQLs based on semiconductor structures have small FEs of less than 0.5 eV, low potential barrier heights of about 0.2–0.5 eV, and large DKs of about 12–14, which does not allow obtaining generation in the optical range. Since QQL has many periods, i.e., it is a fairly extended structure, sufficiently high voltages on the order of tens of volts, low temperatures, or even superconducting metals should be used. The mean free path length (MFPL) of electrons $\lambda_e$ for metals (at the Fermi level) at room temperature is on the order of tens of nanometers (42 nm for copper, 13.3 nm for lead, and 12 nm for beryllium). Estimates using a simple formula $\lambda_e = 1/\left(n_e \pi r_i^2\right)$ based on the concentration ne and the ionic radius $r_i$ [24] yield MFPL values that are an order of magnitude smaller. Next, we consider two metals: copper and lead. The former has high conductivity and a large DOS, while the latter is superconducting at temperatures below 7 K. For copper, we have an electron concentration $n_e = 8.47 \cdot 10^{28}$ m$^{-3}$, FE $E_F$=7 eV, and work function (WF) W=4.4 eV. According to [24], the EM of electrons for copper in CZ is m*=1.5me. However, the FE is consistent with EM $m^*=m_e$, since copper has one unpaired electron that is donated to the crystal lattice. For lead, according to [24], $n_e = 1.15 \cdot 10^{28}$ m$^{-3}$ and m*=2.1$m_e$. There is data that $n_e$=3.3·10$^{28}$ m$^{-3}$. This corresponds to the fact that the atom donates one electron to the lattice. For lead, FE EF=9.45 eV and PV W=4.2 eV. However, one gram-mole of lead is equal to 207.2 g and contains $6.022 \cdot 10^{23}$ atoms. In 1 $m^3$, taking density into account, there are $3.3446 \times 10^{28}$ atoms. Lead has 4 valence paired electrons. If it donates all of them to the crystal lattice, then $n_e = 13.378 \times 10^{28}\,m^{-3}$. In this case, $m^* = 1.024\,m_e$. Further calculations are performed for $m^* = m_e$. Possible metals for constructing QWs and electrodes may include Ag, Nb, and Be. The metal layers serve as QWs, while the cathode and anode serve as electrodes. If the QWs are free (without an external potential being fixed on them), the level of their bottom is determined by the anode voltage. However, in a MIM structure, it is possible to control the level of the QWs by using them as grids and applying specified voltages to them. This opens up additional possibilities for controlling the QP $V(x)$ profile. Applying voltage to QWs allows you to regulate the levels of their bottom, and using additional electrodes in the barriers allows you to control their height. In semiconductor heterostructures, the bottom of the QW has a slope due to the penetration of the field into them. The Debye penetration depth can significantly exceed the size of the wells. In MIM structures, the bottom is flat, since the penetration depth is on the order of an atomic layer, and the QP wells are constant.

The modeling of the structure depends on a number of factors. The first is the absence or presence of a thermal field. If $T$=0, all levels on the electrodes are occupied, and tunneling onto them is impossible. Tunneling is possible only onto levels in QWs or onto the anode with a transition to its Fermi level (FL). In a thermal field, there are always electrons in the tail of the Fermi–Dirac distribution that have freed the levels on the electrodes, so tunneling onto these levels is possible. In a thermal field, level thermalization occurs. The second is the number of quantum wells and the length of the structure. This is especially important at finite temperature. In this case, the parameter $l/\lambda_e$ is important – the ratio of the structure's length to the electron mean free path. When it is significantly less than one, transport throughout the structure is ballistic, and SE can be applied to it. When it is significantly greater than one, transport is diffusive. In this case, SE can only be applied to tunneling through individual barriers if they are narrow. The third is the configuration of the quantum potential. It can be carried out assuming that the voltage drop between the cathode and the anode occurs at the barriers, while the potentials in the wells are constant (Fig. 1). It is possible to apply constant, predetermined voltages to the wells (wells in the form of grids). In particular, configurations of a triode and a tetrode are possible [25]. It is also possible to introduce electrodes with predetermined voltages into the barriers (Fig. 2). This allows one to configure any sufficiently complex potential profile $V(x)$. The fourth parameter is the length of the electrodes (anode and cathode). If it is large, the standard tunneling theory can be used, with the electron wave incident on the barrier, its reflection, and transmission. If it is small, the anode and cathode are also QWs, and tunneling from the cathode is then determined by lifetimes [26].

From the entire variety of heterostructures, we will consider two in detail. The first consists of n+1 identical dielectric layers and n metallic layers between them (i.e., n QWs), with the potential $U_a$ applied between the cathode and the anode. This means that the voltage drops on $\Delta U = U_a/(n+1)$ across each barrier and the quantum potential of each well decreases by $\Delta V = eU_a/(n+1)$ (Fig. 1). For an operating voltage of 10 V and four QWs, we obtain $\Delta V = 2$ eV. For an operating voltage of 100 V and 40 QWs, we obtain generation at the same frequency, but the population of only the first four QWs directly depends on tunneling from the cathode. The population of the remaining QWs is determined by transitions from higher levels. The second structure consists of a barrier, a wide QW, a barrier, and an anode, as shown in Fig. 2. If both barriers are made of dielectric material, the second barrier will be significantly lower than the first, since the first has a height from the bottom of the CZ cathode plus the WF, while the height of the second is determined from the bottom of the QW and equals the WF. For simplicity, we will assume that $U_a=E_F$. Then the bottom of the QW is at the level of the bottom of the CZ cathode, as well as the FL anode. In order to equalize the height of both barriers, we introduce a metal electrode 4 (Fig. 2) connected to the cathode. Otherwise, there will be a large tunneling current in the structure for electron energies above the first level, which reduces the efficiency. The structure can be cascaded, i.e., the number of QWs can be increased by taking the number of periods $N>1$. This cascading leads to the fact that the levels expand into $N$ close sublevels. This increases the transparency area of the level where $D \approx 1$. But in the non-resonant region, transparency decreases exponentially. Fig. 3. This can be explained by the fact that for non-resonant electrons, the width of the total barrier increases $N$ times. Such a structure can have high efficiency, since non-resonant tunneling can be neglected.

If the electrodes (anode and cathode) are made to be extended, with a significant exceeding of the electrode spacing $\lambda_e$, then in this case the density of electrons per unit of volume that impinge on the barrier per second, with velocities in the range from $v_e$ to $v_e = v_e + dv_e$, is given by the formula:

$$dn_x = \frac{m_e^2 k_B T}{2\pi^2 \hbar^3} \ln\left(1+\exp\left(\frac{E_F - E}{k_B T}\right)\right) dv_x . \tag{1}$$

This is the number of electrons per unit volume in the velocity range from $v_x$ to $v_x + dv_x$, obtained by averaging over transverse motions. At zero temperature, one should take $dn_x = m^*(E_F - E)dE/(2\pi^2\hbar^3 v_x)$. At FL, the density is zero and increases to a maximum at the bottom of CZ. If the electrode dimensions are finite and smaller than λe, standing electron waves in them form levels, i.e., the electrodes are quantum wells. In this case, the emission should be considered using the lifetimes of the levels [26]. Thus, we solve a one-dimensional quantum problem, but we take into account the three-dimensional nature of the motion. This means that the one-dimensional levels in QWs are bands that depend on the transverse wave numbers $k_y$ and $k_z$ [26]. We model the barriers as a dielectric with a wide band gap. Electrons moving in the CZ of the metal are screened by the positive lattice ions. Electrons that tunnel and move in the CZ of the dielectric are not screened (there are no free electrons and lattice ions in the dielectric; all atoms are neutral). Therefore, the movement of electrons through a dielectric barrier occurs as in a vacuum, with the only difference being that the field is weakened by a factor of $\varepsilon$. Diamond is the best dielectric for this purpose. Thin nano-diamond films can be deposited using CVD technology [27,28] or by magnetron sputtering from low-pressure plasma. The band gap of diamond is 5.45 eV. With 88 % $sp^3$ hybridization in CVD-grown diamond, the band gap is slightly less than 5 eV. At zero temperature, there are no electrons in the diamond core. The EM of electrons in CZ diamond is $m^*=0.48m_e$ [29]. The DP of diamond 5.6 is quite low, which leads to relatively high barriers. Diamond is the best conductor of heat. There are technologies for producing diamond films with a thickness of several mn with the formation of conductive graphene layers on them [30]. Beryllium oxide BeO is also a good dielectric. Beryllium itself has a maximum FE of 14.6 eV and can also be used to create QWs and electrodes. The field penetrates completely into dielectric barriers. The field practically does not penetrate into metal electrodes: the Debye penetration (shielding) depth $L_D = \sqrt{\varepsilon_0 \varepsilon k_B T / \left(n_e e^2\right)}$ in metals is of the order of an atomic layer at concentrations of $n_e$~$10^{28}$ $m^{-3}$, even at low temperatures of the order of 1 K. This is the advantage of modeling such structures over semiconductor ones: the wells have a flat bottom without a slope, and the slope of the barriers is determined quite simply: the full drop in the anode voltage $U_a$ occurs at the barriers. For example, a potential $V_a(x) = -eU_a x t_{b1}/t_b$ acts on the first barrier, where tb1 is its size, and the total length of the barriers is $t_b = t_{b1} + t_{b2} + ... + t_{b(N+1)}$. Partial field penetration into semiconductor QWs with weak doping requires solving Poisson's equation, which complicates the problem. Such penetration leads to low-threshold emission for conventional diode structures [31]. We strictly define the barrier profile between two pits or between an electrode and a pit by adding the potential of an infinite series of images to the decaying potential $V_a(x)$ caused by the anodic voltage, in accordance with the works [25, 31–36].

$$V_0(x) = \frac{W}{\varepsilon} - \frac{W}{\varepsilon}\delta\left\{\frac{1}{x+\delta} - \frac{1}{2d} + \right. \\ \left. + \frac{2x^2}{d(d-x+\delta)(d+x)} + \frac{2x^2}{d^3}\sum_{n=2}^{\infty}\frac{1}{n^3 - n(x/d)^2}\right\}. \quad (2)$$

The total potential is $V(x)=V_0(x)+V_a(x)$. The potential (2) can be approximated by the function

$$V_0(x)=\frac{W}{\varepsilon}\left(1-\frac{\alpha}{d}\right)\left(1-\frac{\delta d}{[x+\delta(1-x/d)](d-x+x\delta/d)}\right).$$

It also satisfies the condition $V_0(0)=V_0(d)=0$, and at the midpoint $V_0(d/2)$ it coincides with (2). From the condition that the value of the series (2) is equal to the value $V_0(d/2)=W(1-\alpha/d)(1-4(\delta/d)(1-2\delta/d))/\varepsilon$ at $\delta/d<<1$, we obtain $\alpha=2.26\delta(1-2.36\delta/d)$. The potential (2) is derived from the electrostatic Green's function of an electron changing its position x between two electrodes with a distance $d$ between them and a dielectric permittivity $\varepsilon$ between them. It is calculated as the work against the image forces when the electron moves from the cathode to the anode. In this case, the singular part of the Green's function is excluded (the electron does not act on itself), and the image forces do not act at a distance $\delta$ from the electrodes, on the order of the interatomic distance. At the same time, we associate this distance with the WF according to the formula $\delta=1/(16\pi\varepsilon_0 W)$. For W=4.4 eV, we have a size of $\delta$=0.08 nm. It is of the order of the covalent radius of the lattice atom (or the ionic radius), so for $d$~1 nm and more we have $\delta/d<<1$. Thus, the reference for $x$ is not the centers of the outer layer of atoms in the crystal lattice of the electrodes, but a plane shifted by δ, which eliminates divergences in V. For a diode with a barrier (2), $V_0(0)$ = 0 at the cathode and $V_0(d)$ = 0 at the anode ($d$ is the size of the barrier). In this case, the energy is referenced from FL. It is convenient to reference the energy from the bottom of the CZ cathode. Then EF should be added to (2). Further, with a large $U_a$, we will also measure energy from the FL of the anode. In this case, the energy is always positive. It should be noted that the transparency of barrier (2) is significantly greater than that of a rectangular barrier with the same width and height. The barrier profile approaches a rectangular shape at large distances d as it increases. At small $d$, function (2) results in a reduction of the barrier height by a relative amount due to the influence of the electrodes, and the barrier profile approaches a triangular shape. The reduction in height is also provided by function $V_a(x,U_a)$ due to the Schottky effect. At small $d$ and high voltages, the barrier disappears relative to FL and becomes close to triangular [31–36]. For the feeding function (1) of the structure, the thermofield current density through it is given by the formula.

$$J=\frac{em_e k_B T}{2\pi^2\hbar^3}\int_0^\infty D(E,U_a)\ln\left(\frac{1+\exp((E_F-E)/(k_B T))}{1+\exp((E_F-E-eU_a)/(k_B T))}\right)dE. \qquad (3)$$

It is assumed that there is bidirectional tunneling of electrons of all energies, measured from the bottom of the CZ cathode. Accordingly, the transparency $D$ of the structure is the same in both directions, but the number of electrons arriving from the left and right is different. The transparency should be calculated based on the SE solution, taking into account the EMs in the barriers. At $U_a$=0, the current according to (3) is absent. At high voltage, the exponential in the denominator becomes zero, i.e., only one-way tunneling occurs. When $T$=0 and $E > E_F - eU_a$ the logarithm takes the form $(E_F - E)/(k_B T)$. In this case, only one-way tunneling occurs. When $eU_a > E_F$, one-way tunneling occurs for all positive energies $E < E_F$. When calculating integral (3), it is sufficient to take the upper limit to be of the order of two or three EF for the temperatures used. If the EMs are different, the transparency should be calculated taking them into account.

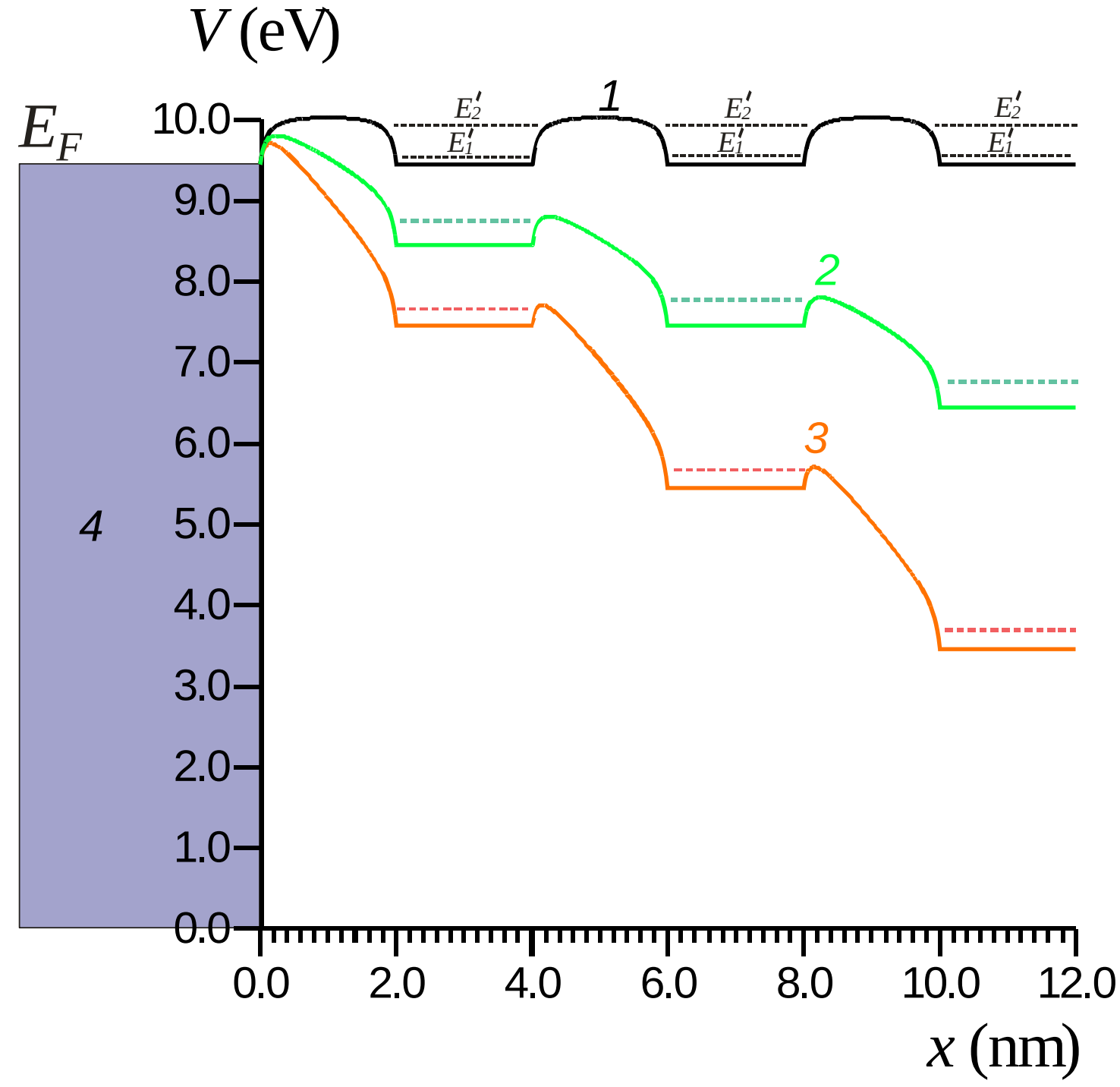

Fig. 1. The profile of the quantum potential $V(x)$ at several wells $d_w$=2 nm and barrier sizes $d_b$=2 nm for PB W=4.2 eV, FE $E_F$=9.45 eV without anode voltage (curve 1), with a 1 V anode voltage drop at each barrier (2) and 2 V on each barrier (3). 4 – CZ of cathode

If the chipboard is comparable to the length of the structure $l$, transparency $D(E,U_a)$ should be replaced with $D(E,U_a)\lambda_e/(\lambda_e + l)$ [37]. In this case, tunneling is no longer fully ballistic, but partially dissipative.

Figure 1 shows the periodic and biased V(x) profiles with pits and barriers of 2 nm in size. In the unbiased anode voltage profile, with a pit depth of 0.55026 eV, there can be two energy levels in it. For n periods, these levels split into n closely spaced levels. This is also evident in the model problem of tunneling through periodic barriers, as shown in Fig. 3. Such periodic barriers can be obtained by applying an anode voltage to the QWs and constructing the barriers as shown in Fig. 2, connecting the metal parts of the barriers to the cathode. Then the height of all barriers is the same, the depth of the pits is the same, and to obtain identical reflections from the barriers, the width of the first barrier should be matched with the others. At voltage $U_a$, the heights and widths of the right barriers decrease, and a single level may be present in the pit at the top of the right barrier. It is populated during tunneling from the cathode, and for pits 2, 3, 4 ... and during tunneling from neighboring levels. Tunneling from the cathode to the level of the fifth pit, curve 3, is impossible (it is negative and not shown). If there are five or more pits, the population of the last levels is determined by tunneling from the levels of neighboring pits. The transparency of the left barrier decreases from pit to pit due to the increase in its width, but at the same time, the number of particles approaching the barrier increases (1). The occupancy of the pit levels due to tunneling from the cathode decreases significantly as the pit number increases. This occupancy decreases exponentially with the width of the barrier, so the simplest solution is to set the occupancy of the first pit level and determine the occupancy of subsequent levels based on transitions to them from the higher levels. Cascade transitions from adjacent levels for profile 3 generate optical quanta with $\hbar\omega$=2 eV. The quanta for curve 2 lie in the IR range. Thus, the structure in Fig. 1, when shifted by a period of about 2 eV, generates optical quanta. The large FE of the cathode makes it possible to populate the first n=$E_F/\Delta V$ levels, where for curve 3 in Fig. 1, $\Delta V$=2 eV. At zero temperature, the FL of the cathode coincides with FE. RL in the wells is lower, and all occupied levels are located below the bottom of the wells corresponding to FE. The levels that arise in the wells at $U_a$=0 are metastable (electrons from them can move to the anode or to the cathode), see Fig. 1, curve 1. However, they do not contain electrons at zero temperature. All electrons are in CZ 4, which lies below. Metastable levels at $U_a$>0 may contain electrons that have passed through the barrier. Non-resonant electrons with energy $E_F - 2 < E < E_F$ tunnel through the left barrier for a potential of 3 from the cathode to the anode. At lower energies, the barrier becomes wide and practically opaque. At the same final temperature of the electrodes, the total tunnel current at $U_a = 0$ is zero (due to reverse tunneling). At the final temperature, some levels are freed by high-energy electrons in the tail of the Fermi–Dirac distribution; there are always free levels, and tunneling in both directions is possible. In formula (1), for the cathode, the chemical potential is approximately equal to FE: $\mu_{cc} = E_F$. However, for the anode, one should take $\mu_{ca} = E_F - eU_a$, which makes the density (1) for tunneling from the anode exponentially smaller. We will assume that when $U_a$ > EF, there is no thermal electron tunneling from the anode. In any case, this effect is very small. At large $U_a$, which is typical for QQL, it can be neglected. It should be noted that the one-dimensional model yields discrete levels in QWs. The electrodes contain zones: due to the high concentration of carriers and their three-dimensional motion, the distances between the levels in the CZ are negligibly small. In reality, the levels in QWs are split. Considering the 1D model of a QW and the 2D transverse free motion of electrons, their wave function can be represented as the product of the wave functions of the discrete and continuous spectra. Therefore, each level in a QW in three-dimensional *k*-space represents a surface [26]. One can speak of the population of levels. The complex energy of a level $E_n = E'_n - iE''_n$ means its finite lifetime $\tau_n = \hbar/(2E''_n)$ [26]. From the perspective of calculating the transparency D(E), the value $1/\tau_n$ is related to the broadening of peaks $D(E'_n)=1$ around the energy $E'_n$ shown in Fig. 3. Calculations show that high levels have shorter lifetimes, i.e., their contribution to *D* is greater. However, the density (1) at zero temperature for high energies decreases and becomes zero for FE. If a level is near FE, its population from the cathode may be small. The population of the lower levels from the cathode is small. However, the population of the levels is determined by all transitions to them and their lifetimes. Formally, when considering the tunneling process and describing the electron's motion with a single wave function for the entire region, the levels should be considered as existing for the entire $V(x)$ profile, i.e., all the wells influence a specific level. This model assumes that the MFPL is significantly larger than the length of the structure. The wave function is determined through tunneling through the barriers. However, with wide barriers, the influence of the QWs on each other is small, and the wells can be considered separately. Furthermore, in a long structure, there may be diffusive (dissipative) tunneling, i.e., some of the electrons are scattered. Also, some electrons from short-lived levels may tunnel further with the emission of a quantum. Diffusive transport and quantum emission are not described as a wave process using the stationary wave function of the Schrödinger equation (SE), which assumes energy conservation. The emission of a photon, i.e., a sudden change in the electron's energy, means the collapse of the wave function. Therefore, tunneling can only be considered prior to such emission.

For example, we can consider electron tunneling only to a certain level if, subsequently, it emits a quantum from that level. In this case, the occupancy of the levels should be determined taking into account tunneling and the probabilities of quantum emission. With a large number of levels in each well, this process is quite complex. It is convenient to consider QWs with one, two, and three levels, assuming that the barriers are wide and the levels in the wells are independent. This significantly simplifies the calculation. Furthermore, this allows us to consider ballistic tunneling if the size of individual barriers is significantly smaller than the MFPL, while the overall size of the structure may exceed the MFPL. There are always electrons on the cathode whose energy does not coincide with any of the levels. If the length of the structure is significantly smaller than the DSB, such electrons tunnel to the turning point near the anode. Formally, their energy does not change. After the turning point, they scatter on the anode's DSB and enter its FL, and then go into the voltage source. Energy $eU_a - E_F + E$ is released at the anode. Only those electrons for which $E > E_F - eU_a$ can tunnel to the anode. After the turning point to the anode, these electrons move in a classically accelerated manner and scatter on the MFPL. This scattering can occur in bursts with the emission of phonons, or once with the emission of a photon. It should be noted that for accelerated high-energy electrons, the MFPL on the anode can change significantly compared to the DSB on the FL of the anode. If the length of the structure l is of the order of or greater than the MFPL, the tunneling coefficient for such electrons can be taken from $\tilde{D} = D(E)\lambda_g/(l+\lambda_g)$ [37]. Such electrons lose energy along the entire structure, and the integral current decreases. If $l << \lambda_g$ all the electrons emitted from the cathode reach the anode ballistically.

The number of levels in a QW is determined by its width, and for a wide well it is proportional to $t_w$. Let a quantum particle traverse the entire potential structure $V(x)$ ballistically without losing energy, i.e., at the average speed of its fall onto the barrier [38,39]. If its energy coincides with a level, it first transitions to it. If the lifetime of the level is significantly shorter than the time required to traverse the structure, the particle may transition from the level to the same or a lower level, but with the emission of a quantum. In the case of several wells, this may be accompanied by a transition into a neighboring well. The transition is accompanied by the emission of a quantum (photon) or phonons. The same applies to other levels. The particle's momentum inside the barrier is imaginary; however, tunneling is an energy-conserving process described by the stationary SE. Therefore, it is quite reasonable to assume that a particle at level $E'_n$ has a velocity $v_n = \sqrt{2E'_n/m_e}$ and to define the time it takes for the particle to cross the barrier into the neighboring well as $\tau_{bn} = t_b/v_n$. Here, $t_b$ is the width of the barrier between the two turning points at level $E'_n$. The transition is most likely if $\tau_n = \hbar/(2E''_n)$. Such transitions are not described by a wave function for the entire potential $V(x)$. Modeling of tunneling processes based on nonstationary SE shows that, indeed, transient tunneling processes are determined by the velocities of particles having different energies [20,38,39]. Wave functions can be introduced for a finite number of wells before such transitions. We will look at these processes below. If there are several n equidistantly located such levels, coherent photon generation is possible. Fig. 4. In this case, it is necessary to consider the population of levels and describe the transition current through the electron lifetime at level [26]. At the same time, from lower levels, if they exist and if their lifetimes are greater than the transit time of the entire structure, the electron moves to the anode, scattering on the MFPL. We assume that the electrodes are quite extended, so scattering in them always occurs. When moving to the FL of the anode, the electron releases an energy quantum and/or phonons. This is a non-coherent process and is not described by a wave function. For conduction electrons (with energy at the Fermi level) at zero temperature $\lambda_e \to \infty$. However, for electrons accelerated by an anode, the mean free path depends on their energy and must be finite. Such an electron excites an atom, releasing phonons. The atom may emit a photon or undergo a non-radiative transition, releasing phonons. Thus, to create coherent radiation, it is necessary to synthesize a structure with equidistant levels and consistent level lifetimes. Typically, isolated QWs and their levels are considered [1]. To analyze the population of levels, thermal relaxation should be considered. The goal of synthesizing QQL structures is to create an equidistant level structure. In this case, MIMs are convenient because they have many parameters: the widths of the barriers and pits, as well as the heights of the barriers and the depths of the pits. The heights of the barriers can be varied by using dielectrics with different dielectric constants. The depths of the pits can be changed by applying specified potentials to their electrodes (grids) [25].

## 2. QUANTUM WELLS AND METASTABLE LEVELS

The solution for the simplest QW of size $t$ with barriers of height $V_1$ and $V_2$ that are infinite in width is given by the formula in [40].

$$k_0 t = n\pi - \arcsin(k_0 t_1) - \arcsin(k_0 t_2), \tag{4}$$

in which $t_1 = \hbar/\sqrt{2m_e V_1}$, $t_2 = \hbar/\sqrt{2m_e V_2}$, $k_0 = \sqrt{2m_e E}/\hbar$, $n$=1,2,.... If necessary, we will replace the electron mass with EM. These are stable levels, and their number strongly depends on the width $t$ of the QW. Another equivalent formula (4) can be given [26]

$$\tan(k_0 t) = -\frac{k_0/\kappa_1 + k_0/\kappa_2}{1 - k_0^2/(\kappa_1 \kappa_2)},$$

or in the form $k_0 t = n\pi - \arctan(k_0/\kappa_1) - \arctan(k_0/\kappa_2)$ where $k_{1,2} = \sqrt{2m_e(E - V_{1,2})}/\hbar = i\kappa_{1,2}$. . Assuming that the barriers have the same height $V$, the well is wide, and considering that the upper level is located near the top of the barrier $V$, we obtain for the number of levels in the well $n = 1 + t\sqrt{2m^* V}/(\pi\hbar)$, i.e., in a wide well, their number is proportional to the width $t$ and the square root of the barrier height. In such a well, the levels are almost equidistant, and the accuracy of their uniform spacing increases with the increase in width. For a well with a width of 6 nm and a barrier height of 10 eV, this formula yields 32 levels. For a 1 nm well, there will be 6 levels. For a 0.34 nm well with a depth of 13.7 eV (graphene), there will be two levels [26]. For very narrow wells, an accurate formula should be used. Narrow wells are convenient for QQL. If they have one level, it is located near the top of the lowest right barrier [40]. However, if there is a single wide QW surrounded by finite barriers, the lifetimes of its nearly equidistant levels drop significantly from the upper level to the lower level if the right barrier has the form shown in Fig. 2. The population of the levels depends strongly on the profile of the quantum potential, the cathode material, and the temperature, i.e., the presence or absence of a thermal field. The lifetimes are determined by the QP profile and correspond to the probabilities of spontaneous emission. In this case, a transition is possible to any of the lower levels with the emission of a quantum, including the FL of the anode. For the transition to the anode, the barrier transit time must be significantly shorter than the level lifetime. The barrier transit time should be defined as the distance between two turning points divided by the velocity corresponding to the level energy. The absence of a thermal field means an infinite DSB and the absence of stimulated emission. At a finite temperature *T*, in a state of thermodynamic equilibrium, the Einstein relation holds

$$\frac{A_{mk}}{B_{mk}}\frac{\pi^2 c^3}{\hbar\omega_{mk}^3}=1, \qquad (5)$$

where $A_{mk}$ characterizes spontaneous emission, and $B_{mk}$ characterizes stimulated emission and absorption. According to Einstein, the probability of transition from state m to state *k* is given by $P_{mk}=A_{mk}N_m+B_{mk}N_m\rho(\omega_{mk},T)$. Here, the thermal radiation density is $\rho(\omega,T)=\hbar\omega^3 f_{BE}(\omega,T)/(\pi^2c^3)$, and $f_{BE}=[\exp(\omega\hbar/(k_BT))-1]^{-1}$ is the Bose–Einstein function. The quantity $N_m$ is the number of electrons per unit area. Accordingly, the dimension of $A_{mn}$ is m$^2$/s, and $P_{mk}^0=A_{mk}N_m$ is the transition probability in the absence of a field. It is also necessary to take into account the transition probability $P_{cm}$ from the cathode to level *m* and from level *m* to the anode $P_{m0}$. The latter probability can be taken in the form $P_{m0}=d_{m0}/v_m$. Here $v_m=\sqrt{2E_m'/m^*}$, $d_{m0}$ is the distance between two turning points in the barrier corresponding to the level. The current density from the cathode for energies in the vicinity of the level at high voltage, according to (3), can be taken in the form

$$J_m=\frac{em_ek_BT}{4\pi^2\hbar^3}\ln\left(1+\exp\left(\left(E_F-E_m'\right)/\left(k_BT\right)\right)\right)E_m''. \qquad (6)$$

Here, we assumed that the spectral line *D* has a triangular transparency profile and is located in the region $E_m'-E_m''/2<E<E_m'-E_m''/2$. Accordingly, $P_{cm}=J_mS/e$ is a probability of transition to the level per unit time for the cathode area *S*. $J_m=em_e(E_F-E_m')E_m''/(4\pi^2\hbar^3)$ at *T*=0. Now the following relationships hold

$$P_{cm}=\sum_{k=0,k\neq m}^{n}P_{mk}^0\left(1+f_{BE}\left(\omega_{mk},T\right)\right), \qquad (7)$$

moreover, for the probability of spontaneous emission from the level, we have

$$\frac{1}{\tau_m} = \sum_{k=0,k\neq m}^{n} P_{mk}^{0} .$$

Relations (6) can be rewritten in the form

$$P_{cm} = \frac{1}{\tau_m} + \sum_{k=0,k\neq m}^{n} P_{mk}^{0} f_{BE}(\omega_{mk}, T).$$

In the absence of a field $P_{cm} = 1/\tau_m$. The total probability of transition from level $k$ is

$$P_k = \sum_{l=0,l\neq k}^{n} P_{kl} .$$

The current density from the level is

$$J_k = e \sum_{l=0,l\neq k}^{n} P_{kl}(N_k - N_l). \quad (8)$$

Here, $n$ is the number of levels. If there are many levels, obtaining solutions to a large number of equations is quite difficult, since it is necessary to determine the transition probabilities. If there is only one level, the only possible transition is to the anode, which is usually the case in RTD. At the same time $P_{c1} = P_{10} = (1 + f_{BE}(\omega_{10}, T))/\tau_1$. Then, at zero temperature, the current density from the level

$$J = I/S = \frac{2eN_1E_n''}{\hbar} = \frac{em_e}{4\pi^2\hbar^3}(E_F - E_1')E'' . \quad (9)$$

Here, the density $N_1 = m_e(E_F - E_1')/(8\pi^2\hbar^2)$ takes into account spin degeneracy, and $S$ is the cathode emission area. If the barrier between the QW and the anode is very narrow, the probability of direct transition to the anode increases. The occupancy of the levels can be controlled by synthesizing the barrier profile between the QW and the cathode. At zero temperature, transition from the levels to the cathode is impossible. Therefore, the lifetimes of the levels are determined by transitions through the barrier between the QW and the anode. By synthesizing such a barrier, it is possible to control the lifetimes so that the occupancies of the levels are set and constant. This is the main condition of the stationary model. The barrier height can be adjusted by creating electrodes with specified potentials, as shown in Fig. 2. For a narrow QW, tunneling can occur directly to the anode; for an electron energy that coincides with a level, the tunneling is resonant. If the QW width is significantly larger than the MFPL, resonant tunneling to the anode is unlikely; resonant electrons populate the levels, while non-resonant electrons tunnel with dissipation into the FL wells and then tunnel to the anode. In such a structure, by calculating the current from the cathode $J_k$ to level k and determining the tunneling current from it through the population and its lifetime, one can determine the steady-state population $N_k$.

Typically, in the literature, rectangular barriers and QWs are considered [1], and the potential is taken as $V(x) = V_{rec}(x) - eU_a x/d$, where $d$ is the length of the structure, and $V_{rec}$ is a rectangular, in particular, quasiperiodic function. Thus, it is assumed that the field penetrates the barriers and wells equally, and the wells have a sloping bottom. Often, even the potentials of the barriers are described by delta functions [7,23]. QWs are considered separately [1,7], although in ballistic transport through several QWs and barriers, the energy levels for the structure as a whole should be found. Indeed, neighboring QWs interact through a barrier via tunneling, and their total wave function is equal to the superposition of the well functions. For a periodic QP (Fig. 1, curve 1), the levels split. Only when the barrier width is very large can they be considered independent. The interaction for n QWs leads to the splitting of levels into n sublevels. In the case of a solid-state crystal with a quasiperiodic potential, the QWs correspond to atoms, which leads to the formation of bands even in the case of nanoscale samples. Figure 3 shows the calculation of the transparency $D$ from the cathode to the anode for a model problem involving one, two, and four rectangular QWs surrounded by rectangular barriers with dimensions of $t_b$=0.5 nm, $t_w$=6 nm, with $U_a$=10 V and a voltage across the wells (grids) of $U_g$=10 V. This is a model problem that uses rectangular barriers of equal height; however, the results are qualitatively similar for real barriers, as shown in Figure 2.

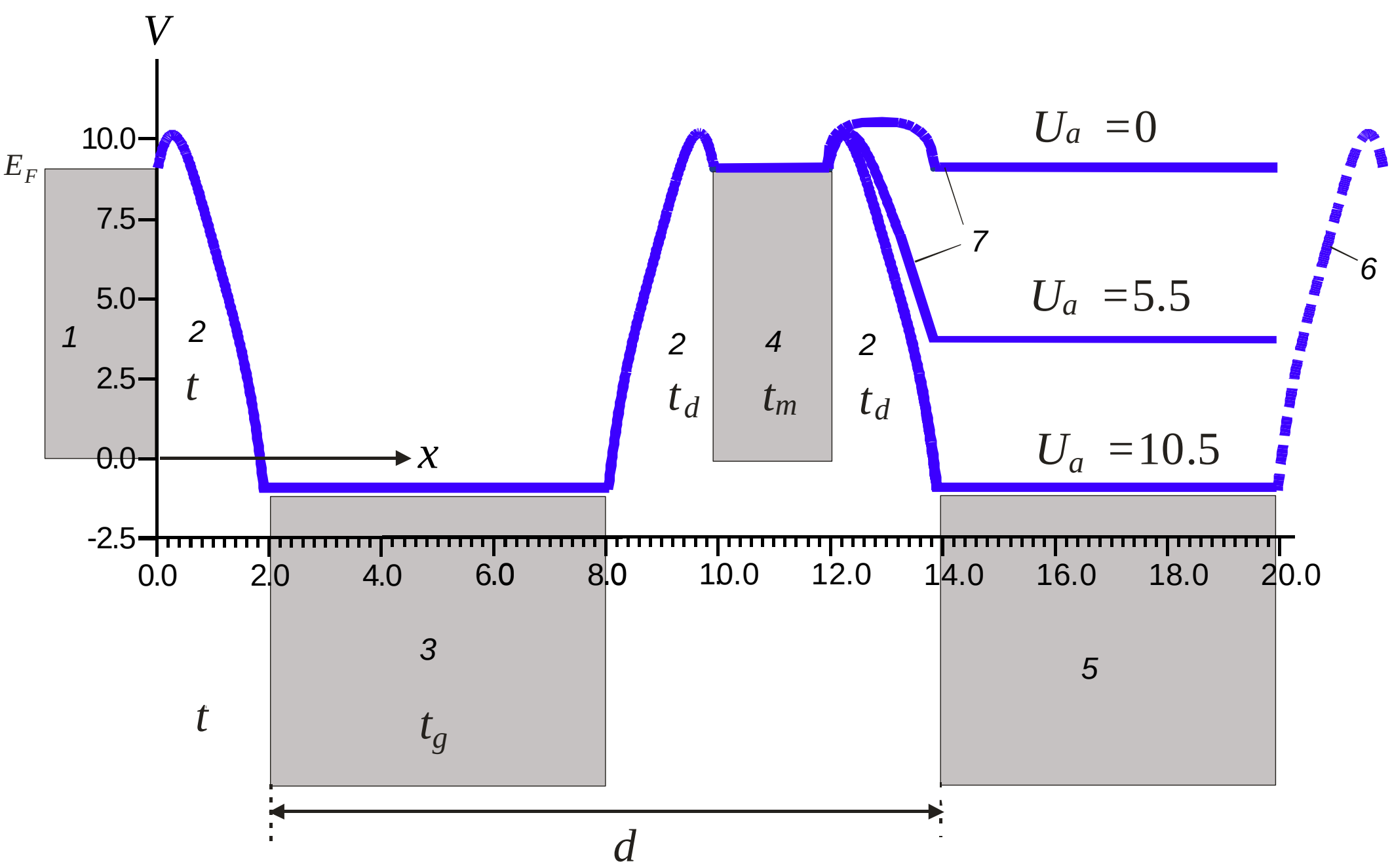

Fig. 2. Distribution of the quantum potential $V(x)$ (eV) in a Pb-diamant heterostructure at a fixed voltage across the wells, depending on the length (nm) for one period. The gray areas indicate the CZ cathode (1), the grids (3), the electrode in the barrier (4), and the anode (5); 2 indicates the barrier regions with dielectric layers. Voltages $U_a$=0, 5.5, and 10 (V) are considered. $E_F$=9.45 eV. The periodic continuation is shown as 6.7 – the potential at the anode in the case of one period.

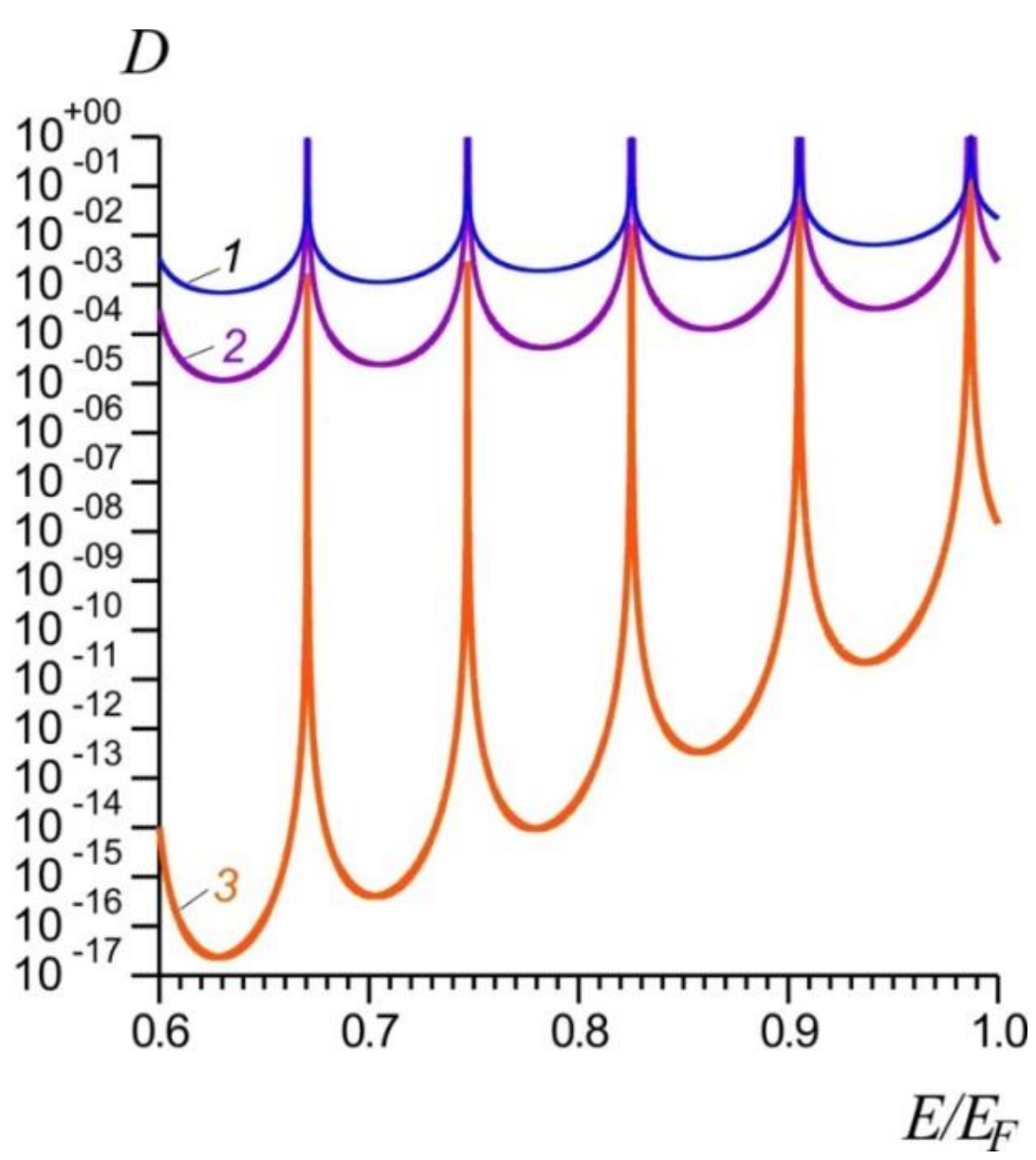

Fig. 3. Tunneling coefficient $D$ in one (curve 1), two (2), and four (3) QWs with a width of 6 nm at $t_b$ = 0.5 nm, $U_a$ = $U_g$ = 10 V

For the structure shown in Fig. 2, all barriers except the first one consist of three parts: a dielectric part, a metallic part, and then another dielectric part. The metallic parts are connected to the cathode, and the QWs (grids) are connected to the anode, i.e., an anode voltage is applied to them. This makes it possible to align all the barriers in height and obtain a quasiperiodic structure with an equidistant spectrum of levels. This is demonstrated by the calculation shown in Fig. 3, which demonstrates an equidistant arrangement of peaks, i.e., the levels are also equidistant. The extremely small width of the barriers is taken for the sake of clarity in this model problem. A larger width leads to very narrow peaks. A detailed analysis of the peak region shows that there is a fine splitting into n maxima, reaching unity. This indicates that the levels are also splitting. However, an increase in the number of periods does not lead to an increase in current: although the levels become broader, the decay of tunneling between them increases, and the current of non-resonant electrons decreases. The same applies to an increase in the width of the wells. For a single QW, an increase in its width leads to an increase in resonant levels. However, the transparency between these levels decreases, as a result of which the integral current saturates as the QW width increases. This occurs when tw~20 nm or more. Increasing the number of periods in the structure makes it possible to increase the width of the levels (reduce their lifetime) and also to reduce the transparency between them, i.e., to reduce the non-resonant current. The short lifetime in a long structure leads to the fact that electrons are forced to move to adjacent levels, emitting an energy quantum. A transition through one or two levels is less likely. Such transitions can lead to the generation of harmonics.

For a well with $N$ levels, we can write the balance equation for level $i$ in terms of the transition times between levels $\tau_{12}$, assuming that external filling $J_{in}/e$ occurs at the upper level, and current $I_{out}$ is lost from the lower level. In the steady-state case $dn_i/dt=0$, and we get $I_{in}=I_{out}=I$. For three levels, we have

$$\frac{dn_3}{dt}=I_{in}+\frac{n_1}{\tau_{13}}+\frac{n_2}{\tau_{21}}-\frac{n_3}{\tau_{31}}-\frac{n_3}{\tau_{32}},$$

$$\frac{dn_2}{dt}=\frac{n_3}{\tau_{32}}+\frac{n_1}{\tau_{12}}-\frac{n_2}{\tau_{21}}-\frac{n_2}{\tau_{23}},$$

$$\frac{dn_1}{dt}=-I_{out}+\frac{n_2}{\tau_{21}}+\frac{n_3}{\tau_{31}}-\frac{n_1}{\tau_{12}}-\frac{n_1}{\tau_{12}},$$

In the steady-state mode, the time derivatives are equal to zero and $I_{in}=I_{out}=I$. Thus, the general equation for the velocity of electrons in subband *i* of a system consisting of *N* levels is as follows.

$$\frac{dn_i}{dt}=\sum_{j=1}^{N}\frac{n_j}{\tau_{li}}-n_i\sum_{j=1}^{N}\frac{1}{\tau_{ij}}+I\left(\delta_{1N}-\delta_{i1}\right)=0.$$

For three-level wells in the low-temperature approximation (in the absence of absorption), population inversion is possible. Accounting for all possible situations leads to more complex equations.

Let us consider the characteristic transit times and the lifetimes of the levels. The structure is shown in Fig. 2 with $t_w$=6 nm, and it has *N*=32 levels in the QW. All levels are populated from the cathode. The Fermi velocity for Pb is $v_F=1.6\cdot 10^6$ m/s. For one period, the length of the structure is 14 nm. An electron with the energy of the upper level traverses it in 8.75 fs. An electron at a low level with an energy of 1 eV traverses it approximately 3 times more slowly, and electrons at intermediate levels traverse it approximately 1.5 times more slowly. For the structure to pass through, the lifetime of the level $\tau_n=\hbar/\left(2E_n''\right)$ must be on the order of or greater than this value. For the upper level, this means $E_n''/E_n'<N\cdot 0.004$ that it is not satisfied even for several periods. The transit time for the middle and lower levels does not increase significantly, but their lifetimes increase greatly. This leads to a situation where the lower levels are freed due to tunneling from them through the entire structure, while transitions from the upper levels to lower levels are possible. To achieve such QQL, metals with a large FE are required, for example, lead or Be. Then the energy interval for a QW with n levels is . For lead with 6 levels in the well, $\hbar\omega\approx 1.5$ eV; for narrow wells with 3 levels, $\hbar\omega\approx 3$ eV. By creating wide wells, it is possible to obtain generation in the IR, THz, and even mm ranges. Wide wells and a large number of periods require low temperatures. It is possible to use superconducting metals. In this case, the distance between the levels must be significantly greater than the energy gap in the superconductor, so that Andreev reflection (interaction) can be neglected.

## 3. DETERMINATION OF METASTABLE ENERGY LEVELS

By dividing the profile $V(x)=V_0(x)-V_a(x,U_a)$ into a large number of points xn, where the function $V_a$ describes the voltage drop across the barriers, we determine the wave number $k_n=\sqrt{2m^*(E-V(x_n))}/\hbar$ and the normalized wave impedance $\rho_n=k_0/k_n$ for each of them. The potential is measured from the bottom of the CZ cathode, so when $U_a > E_F$, the values of $V(x_n)$ may be negative. At the anode. We introduce normalized wave impedances as the ratios of wave functions to their derivatives, or $\rho_0=ik_0\psi/\psi'$ [25]. In the region of the barriers, these quantities and wave numbers may be imaginary. Let an electron tunnel through the structure ballistically, and its energy does not coincide with any level. In this case, it tunnels with significant reflection. In the structure shown in Fig. 1, after the last turning point, it moves quasi-classically, accelerating above the potential, and reaches the anode with energy $eU_a-E_F+E$. In this case, it is described by the wave function and coefficient $D$ only up to the turning point. In the case of conventional diode tunnel structures, the turning point is located near the anode [33–36]. In the case of a large number of periods (QWs) and a large $U_a$, the turning point is located at the beginning of the structure. Within the free path in the anode, the electron relaxes to the anode FL, releasing energy $eU_a-E_F+E$. This energy can be released in quanta through successive scattering, releasing optical quanta and phonons, or immediately through the emission of a single quantum. This is not a wave process. Let's say an electron excites an atom on the anode, releasing phonons. The excited atom transitions to a lower state, emitting a photon. This is incoherent radiation. It is associated with electrons of different energies that are present in the flow approaching the barrier. There are few such non-resonant electrons. At low $U_a$, on the order of fractions of a volt, it can be assumed that scattering on the MFPL leads to phonon excitation and heating. At voltages on the order of tens of volts, UV photon emission is possible. Such high-energy electrons have a different MFPL on the anode. Let us now consider the case when an electron with energy $E$ tunnels without reflection. This is the case of resonant tunneling. Electrons with a certain energy in the flow may be reflected or pass through the barrier and the entire structure. To determine this condition, we will recalculate the wave impedance from the anode to the cathode using the impedance transformation formula [31–36]. For resonant tunneling, the reflection coefficient must become zero: $R(E)=(\rho_0-\rho_{in})/(\rho_0-\rho_{in})=0$. This is the condition for the existence of metastable energy levels $E=E_n=E_n'-iE_n''$ [25]. It is divided into two parts: for the real and imaginary parts of the reflection coefficient. In [25], a similar equation is derived based on the transmission matrix. This method allows for the consideration of arbitrary profiles $V(x)$.

The problem of levels in QW can be analyzed in a different way by considering only outgoing or exponentially decaying electron waves. In particular, by considering a QW of width tw surrounded by two barriers of height $V$, the left one being infinitely wide and the right one having a width of $t_b$, we obtain that the complex levels are determined by the equation $\tan(k_0 t_w + \varphi) = \Phi(k_0, \kappa)$, in which

$$\Phi(k_0,\kappa) = -\frac{k_0}{\kappa}\frac{\left(1-\frac{ik_0}{\kappa}\right)\exp(\kappa t_b)+\left(1+\frac{ik_0}{\kappa}\right)\exp(-\kappa t_b)}{\left(1-\frac{ik_0}{\kappa}\right)\exp(\kappa t_b)-\left(1+\frac{ik_0}{\kappa}\right)\exp(-\kappa t_b)}, \tag{10}$$

$\tan(\varphi) = k_0/\kappa$, $\kappa = \sqrt{2m^*(V-E)}$, $k_0 = \sqrt{2m^* E}$. Here, EM is introduced. In the case of a thick barrier $\Phi(k_0,\kappa) = -k_0(1+2\gamma)/\kappa$, where $\gamma = \gamma' + i\gamma''$ is an exponentially small complex parameter with a real part and $\gamma' = \exp(-2\kappa t_b)(1-k_0^2/\kappa^2)/(1+k_0^2/\kappa^2)$ imaginary part $\gamma'' = 2\exp(-2\kappa t_b)(k_0/\kappa)/(1+k_0^2/\kappa^2)$. The complex levels are found iteratively from the equation with a complex function $\Phi$:

$$E_n = \frac{\hbar^2}{2m_e t_g^2}\left[\arctan\left(\frac{\Phi(k_0,\kappa)-k_0/\kappa}{1+\Phi(k_0,\kappa)}\right)+n\pi\right]^2. \tag{11}$$

To explicitly find $E_n''$, we use the condition $E_n''/E_n' << 1$ (the width of the barrier is large) and the decomposition $k_0 = \sqrt{2m_e E'}(1 - iE''/2E')/\hbar$. Separating the imaginary and real parts and assuming that all parameters depend only on $E_n'$, we obtain explicit solutions in the form $E_n'' = 8\cos^2(X)\gamma'' E_n'/(t_g\kappa)$ where $X = \sqrt{2m_e E'}t_g/\hbar + \arctan(k_0/\kappa)$, $E_n'$ is determined from (11) for $\gamma'' = 0$. In general, equation (11) is solved iteratively for each level. A more difficult task is to find the QW levels surrounded by two barriers of different heights $V_1$ and $V_2$ with dimensions $t_1$ and $t_2$ Fig. 4. The wave number in the QW is $k_0$, in the barriers it is $k_1$ and $k2$, behind the second barrier it is $k_3 = \sqrt{2m^*(E+\Delta V_2)}/\hbar$, and in front of the first barrier it is $k_4 = \sqrt{2m^*(E-\Delta V_1)}/\hbar$. This means that the level after the second barrier is lower than the QW level by $\Delta V_2$, and the level in front of the first barrier is higher by $\Delta V_1$. After transformation through the right second barrier to the well, we will have an input impedance.

$$Z_2 = -i|\rho_2| \frac{\rho_3 - i|\rho_2|\tanh(|k_2|t_2)}{-i|\rho_2| + \rho_3 \tanh(|k_2|t_2)}.$$

With a large barrier thickness $\tanh(|k_2|t_2) = 1 - \delta_0$, and we have $Z_2 = -i|\rho_2|(1+\delta_0) \approx -i|\rho_2|$. The real part of the impedance is small. Similarly, the transformation of the impedance $\rho_4$ by the left barrier to the well will give

$$Z_1 = -i|\rho_1| \frac{\rho_4 - i|\rho_1|\tanh(|k_1|t_1)}{-i|\rho_1| + \rho_4 \tanh(|k_1|t_1)}.$$

We will take the wave function in the well in the form $\psi(x) = A\sin(k_0 x + \varphi)$. On the left boundary of the well $\psi'(0)/\psi(0) = k_0 \cot(\varphi) = -ik_0 Z_1$, on the right boundary of the well $\psi'(t)/\psi(t) = k_0 \cos(k_0 t + \varphi) = ik_0 Z_2$, from which we obtain the equation for determining the energy levels

$$\cot(k_0 t + \arctan(i/Z_1)) = iZ_2. \tag{12}$$

In these formulas for the barriers $k_{1,2} = i\kappa_{1,2}$, $\rho_{1,2} = -ik_0/\kappa_{1,2}$. The electron wave from the structure to the left must decay exponentially. From (12), assuming the barriers are wide and $E''_{n.}/E'_{n.} << 1$, we have $\sqrt{2m^* E'_n}\left(1 - iE''_n/(2E'_n)\right)t/\hbar = \arccos(iZ_2) - \arctan(i/Z_1) + 2(n-1)\pi$. Separating the real and imaginary parts, we obtain

$$\sqrt{E'_n} = \hbar \frac{\mathrm{Re}(\arccos(iZ_2) - \arctan(i/Z_1)) + 2(n-1)\pi}{\sqrt{2m^*}t}, \tag{13}$$

$$E''_n = -2\hbar \frac{\mathrm{Im}(\arccos(iZ_2) - \arctan(i/Z_1))}{t\sqrt{2m^*/E'_n}}. \tag{14}$$

By determining the actual level from equation (13), we substitute it into equation (14) and obtain the solution to the problem. Since the impedances depend on the energy, equation (13) should be solved iteratively. As an initial approximation, we can take the solution to problem (4) for the QW with infinitely wide barriers. The generalization to an arbitrary barrier profile (red dashed curves in Fig. 4) simply involves recalculating the impedances $Z_1$ and $Z_2$ for them.

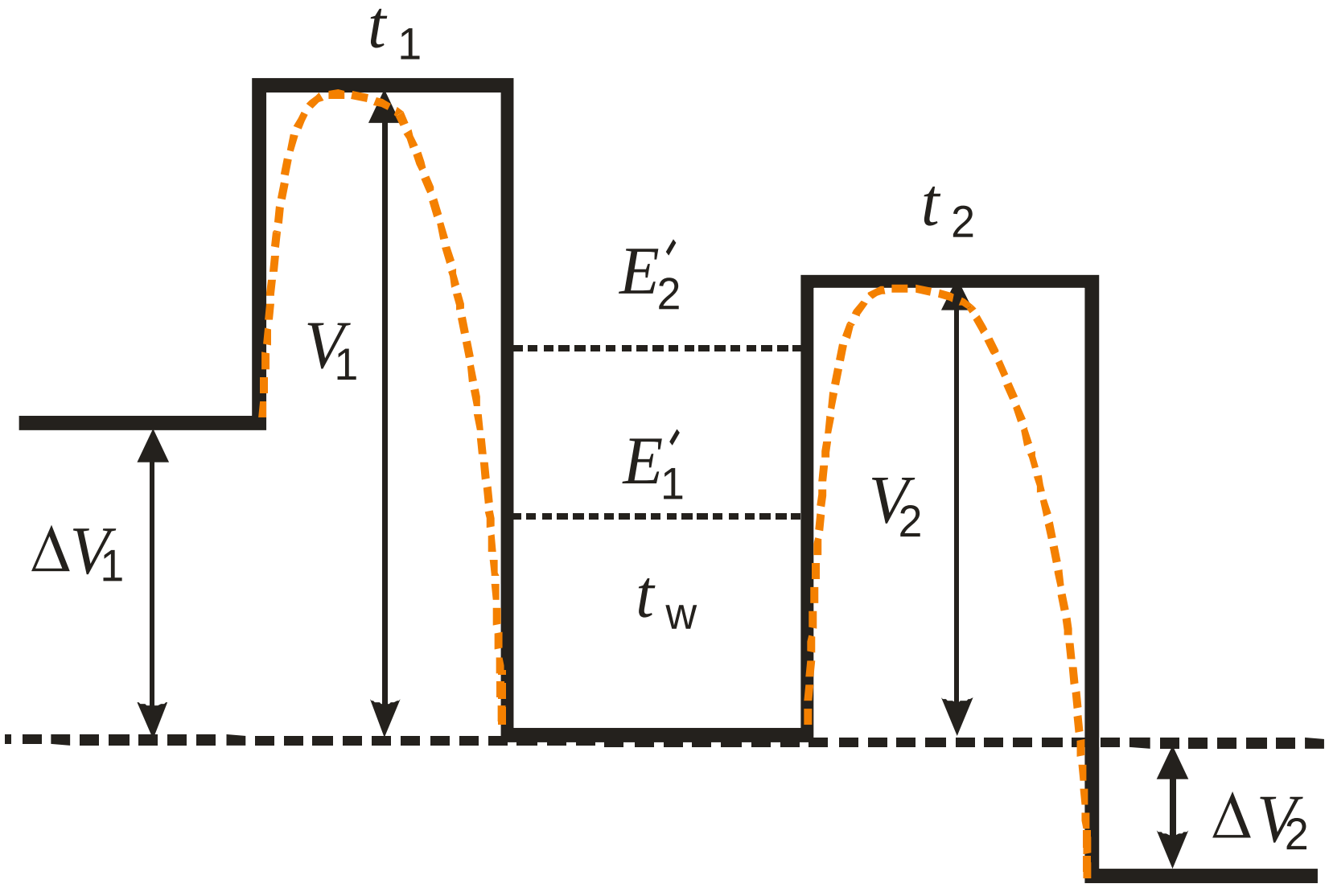

Fig. 4. The structure of the QW potential between a well shifted by a potential $\Delta V_1$, a barrier $V_1$, and a barrier $V_2$, followed by a well shifted by a potential $-\Delta V_2$

Thus, it is possible to construct a structure $V(x)$ without fixed voltages on the QW and with fixed voltages. In the first case, the first QW has its bottom below the cathode's FL by $t_{b1}U_a/t_b$, the second QW has its bottom below the bottom of the first QW by $t_{b2}U_a/t_b$, and so on. Here $t_b=t_{b1}+\ldots+t_{b(N+1)}$, where $N$ is the number of QWs. In the case of identical barriers, each of the wells (including the Fermi level (FL) of the anode) drops by a value of $\Delta V = eU_a/t_b$. The widths of the QWs remain free parameters. They can be used to control the levels and their number. If $eU_a >> E_F$ In a large number of QWs, most of them may have levels to which transitions are impossible directly from the cathode levels. Such wells can be considered isolated. The most likely transitions are possible between adjacent QWs, since distant wells are separated by several barriers. We will consider tunneling between two wells in the next section.

## 4. TUNNELING BETWEEN TWO QWs AND POPULATIONS OF LEVELS

Let the electron on the cathode not be in resonance with the levels of the wells; it is either reflected or tunnels to the anode with a small coefficient $D$. The contribution of such electrons to the total current is insignificant, and the radiation is incoherent. Electrons with energies around $E_n'$ can tunnel to these levels in the first QW. Due to the finite lifetime, the level broadens, so tunneling of electrons from the region $-E_n''/2+E_n' < E < E_n''/2+E_n'$. is possible. In the center of the region, $D$=1; at its edges, the coefficient is $D<1$, but not too small. According to the formula

$$J = \frac{em^*}{2\pi^2\hbar^3} \int_{E_n' - E_n''/2}^{E_n' + E_n''/2} D(E)(E_F - E) dE \ . \quad (15)$$

It is possible to calculate the tunnel current density through the first barrier and determine the surface population of the level $N_m$. Indeed, $J = n_v e v_x = n_v e\sqrt{2E_n'/m^*}$ where

$$dn_x = m^* \tilde{D}(E_n')(E_F - E_n')dE / \left(2\pi^2\hbar^3 v_x\right),$$

$\tilde{D}$ is an easily calculable average tunneling coefficient per level. Now, the occupancy of the level is $n_x = m^* E_n''(E_F - E_n')\tilde{D} / \left(2\pi^2\hbar^3 v_x\right)$. According to (15)

$$J = \frac{em^* \tilde{D}(E_n')}{2\pi^2\hbar^3}\left(E_F E_n'' - (E_n' + E_n''/2)^2/2 + (E_n' - E_n''/2)^2/2\right) \approx \approx \frac{em^* \tilde{D}(E_n')E_n''}{2\pi^2\hbar^3}\left(E_F - 2E_n'\right).$$

According to another formula $J = em^* \tilde{D}(E_n')E_n''(E_F - E_n')/\left(2\pi^2\hbar^3\right)$. A significant difference occurs for the level near FL. In this case, the integration in (15) is incorrect; there are no electrons with energy $E_F + E_n''$, and the more accurate second formula should be used. Per second, electrons per unit area $J/e$ are supplied to the level. Due to the finite lifetime of the level, the tunnel current density from it is determined by formula [25].

$$J = I/S = \frac{2eN_n E_n''}{\hbar}, \quad (16)$$

where $S$ is the emission area of the cathode, $N_n$ $m^{-2}$ is the surface density of electrons at the level. The spin degeneracy factor $g_s = 2$ is taken into account here. Formula (16) corresponds to the definition of tunneling via the lifetime, which originally occurred when the tunneling effect was discovered. By equating the leakage current (16) to the supply current of the level, we obtain the surface occupancy ($m^{-2}$)

$$N_n(E_n') = \frac{m^* g_s \tilde{D}(E_n')(E_F - E_n')}{4\pi^2\hbar^2}. \quad (17)$$

This formula shows that the population density near FL is low. During tunneling from the cathode, the outgoing electron with energy $E$ is replaced by an electron with FL, releasing energy $E_F - E$ (the Nottingham effect of cathode heating). Therefore, there can be no reverse tunneling to this level. The electron may transition with emission to a lower level in the first well, or to a level in another well. If the lifetime of the level is significantly shorter than the transit time of the structure, tunneling directly to the anode is impossible. The simplest approach is to consider narrow wells with one, two, or three levels. In this case, we use a simplification, assuming that the level is created solely by the configuration of the well itself and does not depend on neighboring wells. This is valid for wide barriers. For single-level wells, tunneling into a neighboring well is most likely. If the levels of all wells are equidistant and shifted by $\Delta E$, coherent tunneling with the emission of quanta $\hbar\omega = \Delta E$ is possible. The population of the levels is mainly determined by neighboring wells. The number of emitted quanta is determined by the difference in population densities. The population densities should be determined using the method described above, taking into account all the wells, their levels and lifetimes, as well as tunnel transitions through the barriers between them. This is a rather complex procedure. In the case of two-level wells, it is possible to transition from the cathode to the higher level, followed by the emission of a quantum and a transition from that level to the lower level. It is also possible to transition from the cathode to the lower level of the first well. Subsequently, from the lower level, it is possible to transition to the upper level of the second QW by emitting a quantum, and so on. Here, it is also necessary to determine the population densities of all levels. In the case of three-level wells, it is also possible to transition with the emission of a quantum from the highest level to the second, and from the second to the lowest level. If the lowest level of the first well corresponds to the highest level of the second QW, a tunnel transition to it is possible. The process then repeats. It becomes more complex in the case of multilevel wells. In this case, it is difficult to ensure equidistance of all levels, and the radiation is partially coherent. It is also incoherent in the first two cases if the levels are not equidistant. It is convenient to consider narrow wells with one and two levels, for which the relationships are simpler.

Let us assume that after the quasi-periodic heterostructure there is a wide well, several times the size of the superlattice, with a fixed potential applied to it, and after that there is again a quasi-periodic heterostructure. In fact, this means cascading QQL. The wide well is equivalent to a cathode. Its levels are very closely spaced and play the role of a CZ cathode. The upper level corresponds to the FL. Since electrons are scattered multiple times, the levels overlap, and formula (1) should be used. In principle, it is possible to ensure two or more different distances $\Delta V$ between the levels in each structure or in two different structures, i.e., generation at two or even several frequencies. The advantage of MIM structures is that for two-level wells, their levels can be easily adjusted using small external voltages, as well as by changing the barriers. The task of synthesizing the $V(x)$ profile is inverse: by specifying the arrangement of the levels $(E'_n)$, one should determine the configurations of the wells and barriers.

## 5. PERIODICALLY DISTRIBUTED QWs AND BARRIERS

The simplest case is when there is a quasiperiodic heterostructure with *N* identical wells and *N+1* identical barriers. If the structure has a length *l* greater than $\lambda_e$, then for non-resonant electrons, the transparency is determined as $D(E)\lambda_e/(l+\lambda_e) << 1$.

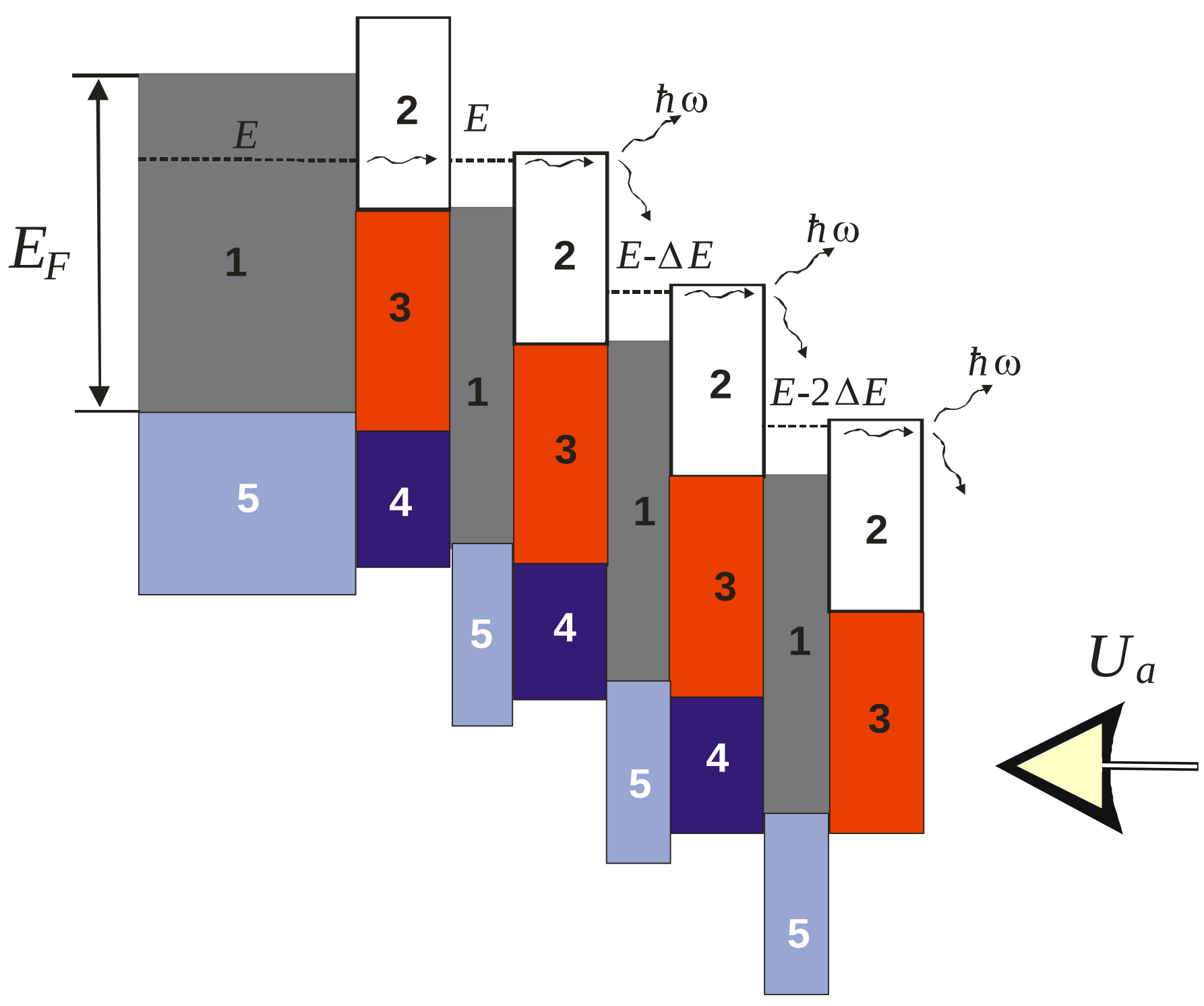

Fig. 5. QQL scheme with a single level in narrow QWs. 1 – CZ of the cathode and QW, 2 – empty CZs of the dielectric barriers, 3 – forbidden bands of the dielectric barriers, 4 – valence bands of the dielectric barriers, 5 – valence bands of the cathode and QW. All wells are shifted by an energy $\Delta E$

These electrons primarily release heat in the form of phonons throughout the structure and do not contribute to radiation. Let the period size be significantly smaller than $\lambda_e$. Then the transport to neighboring wells is ballistic in the form of tunneling. Each well descends by the $\Delta V = eU_a/(N+1)$. We measure energy from the FL of the anode. Then the FL of the cathode has a value $V_F = eU_a + E_F$. Let $eU_a >> E_F$. Then, most of the pits are located below the bottom of the CZ cathode. An energy quantum can be obtained over a wide range of frequencies. Thus, at Ua=100 V and with 50 wells, we get $\Delta E$=2 eV. At the same voltage and with 500 wells, we have 0.2 eV, and at $U_a$=10 V, the quantum is 0.02 eV. This is almost in the THz range. However, the condition introduced above is not met. To achieve this, it is necessary to increase the number of periods, which makes the design of THz QQL more difficult. It is also possible to reduce the voltage, which complicates the calculation. Increasing the length leads to an increase in power.

In principle, FL QQL can be obtained at high voltages. At the same time, the right barriers become very small and narrow, so the width of the pits should be increased to ensure the existence of a single level in them.

Let us consider the synthesis of the structure at $\hbar\omega = 2\,\mathrm{eV}$. Choosing a barrier wide enough at 2 nm, we obtain its excess over FL by 0.7 eV. Let the bottom of the first pit be 2 eV lower than FL. Then this barrier decreases to 0.25 eV and has a height of 2.25 eV relative to the bottom of the pit, and the right barrier has a height of 0.1 eV. In such a QW, there is one level near the right barrier with a short lifetime approximately equal to the transition to the neighboring QW Fig. 1. The diagram of such a QQL is shown in Fig. 5.

## 6. WIDE QWWITH A BARRIER OF $E_F$ HEIGHT

Let us consider a wide QW with width tw and depth EF relative to the FL of the cathode at $T$=0. In this case, the cathode can be considered as an infinitely extended barrier. Moreover, between it and the well, there is a finite barrier of slightly greater height, determined by the cathode's WF. Such a QW is obtained by applying a potential $U_g=U_a=E_F/e$ to the metal layer of width $t_w$. In this case, the bottom of the well is located at the FL of the anode and the bottom of the conducting zone of the cathode. Tunneling from the levels to the cathode is impossible, so their lifetime is determined by tunneling to the anode through the right barrier. For a cathode with a large RV and for barriers with a large DC, their height above the cathode's FL becomes quite small. Such a barrier can be replaced with an infinite rectangular barrier of height EF to the left of the well. The wave function in it has the form $\psi(x) = \psi_0 \exp(x\kappa_0)$, where $\psi(x) = \psi_0 \exp(x\kappa_0)$, $E$ is the energy of the level in the well. We measure the coordinate from the beginning of the well. We take the wave function in the well in the form $\psi = \psi_1 \sin(k_0 x + \varphi)$, $k_0 = \sqrt{2m^* E}/\hbar$. The wave function at the anode is $\psi = \psi_a \exp(ik_0 t - t_w - t_b)$. The wave function in the barrier region consists of evanescent waves in both directions. At each point of the barrier, there is a wave number $k = \sqrt{2m^*(E - V(x))}/\hbar$ and waves $\psi = A^{\pm} \exp(\pm ikx)$ in both directions associated with it. By matching the wave function at the left boundary of the well, we obtain $k_0/\kappa_0 = 1/\sqrt{E_F/E - 1} = \tan(\varphi)$. By transforming the normalized wave impedance $z = k_0/k$ from the anode to the well, we obtain the input impedance $Z_{in} = ik_0\psi(t_w)/\psi'(t_w)$. It is complex, but for a wide barrier it is almost purely imaginary. Such a transformation can be easily obtained for a barrier of arbitrary profile [25,26,31]. The characteristic equation should be taken in the form $ik_0\psi(t_w)/\psi'(t_w) = i\tan(k_0 t_w + \varphi) = -Z_{in}$, $\rho = -ik_0/\kappa_0$. For a rectangular barrier, we have $Z_{in} = \rho[1-\rho\tanh(\kappa t_b)]/[\rho-\tanh(\kappa t_b)]$, $\rho = -ik_0/\kappa_0$. For a wide barrier $Z_{in} = -\rho$. For an infinitely high barrier $Z_{in} = 0$. Here $\kappa_0/k_0 = \sqrt{E_F/E-1}$. For a level $E'_n \approx E_F$ of about FE $\kappa_0/k_0 = \tan(\varphi) = 0$, $\tan(k_0 t+\varphi) = \infty$, $k_0 t + \varphi = \pi/2 + n\pi$, and we get the number of levels $n = t_w\sqrt{2m^* E_F}/(\hbar\pi) + 1/2$. For a level near the bottom $k_0/\kappa_0 \approx 0$, $k_0 t = \pi/2$, and $E'_1 = (\hbar\pi)^2/(8t_w^2 m^*)$. For an arbitrary barrier, we obtain a characteristic equation $\tan(k_0 t+\varphi) = iZ_{in}$. From which complex frequencies are determined. Their real parts can be approximately taken in the form $E'_k \approx kE_F/n$, $k$=1,2,...,$n$. Let this be a zero approximation. The exact iterative formula has the form

$$E_n = \frac{\hbar^2}{2m^* t_w^2}\left[\arctan\left(\frac{iZ_{in} - k_0/\kappa_0}{1 + iZ_{in}k_0/\kappa_0}\right) + n\pi\right]^2 . \tag{18}$$

The levels are not exactly equidistant, but the spread is less than their width. In a wide well, on the order of several nanometers, the number of levels is large — on the order of dozens — and such a structure can generate radiation in the IR and THz ranges. If there are several identical QWs, the levels become broader, and tunneling between the levels is practically absent. With a QW width of 2 nm, there are 6 levels, i.e., generation is possible in the optical range. However, in this case, the levels are no longer nearly equidistant, and the radiation is not entirely coherent.

## 7. LEVEL POPULATIONS DURING THERMALIZATION

Let there be $n$ equidistant levels in one or more QWs. We take the FL of the anode as the zero level. Such a model is possible with a periodic QP, Fig. 1, when a level consisting of n sublevels splits into $n$ separate levels shifted by $k\Delta V$ as a result of applying an anode voltage, as well as for a wide single QW, Fig. 2. Let $N_k$ be the population of level k. Let $P_{kl} = A_{kl}N_k + B_{kl}N_k\rho(\omega_{kl}, T)$ be the transition probabilities between levels. For equidistant levels $\omega_{kl} = \omega = \Delta V/\hbar$. There is a full probability of moving from level $k$

$$P_k = \sum_{l=0, l\neq k}^{n} P_{kl} = 1, \tag{19}$$

$$P_{kl} = A_{kl}N_k\left(1 + \frac{\hbar\omega_{mn}^3}{\pi^2 c^3}\rho(\omega_{kl}, T)\right). \tag{20}$$

Let transitions from the cathode to levels $n$, $n$-$1$, ..., $n$-$k$ be possible. They are determined by the resonant current density (3) from the cathode. We write this density in the form

$$J_g = \frac{em_e k_B T}{4\pi^2\hbar^3}\sum_{l=0}^{k}\ln\left(1 + \exp\left((E_F - E'_{n+l-k})/(k_B T)\right)\right)E''_{n+l-k} . \tag{21}$$

We believe that it is precisely this that determines the generation. It can be represented as $J_g = J_{gn} + J_{g(n-1)} + ... + J_{g(n-k)}$. The non-resonant current density from the cathode is mainly provided by the electrons that have passed through, above the maximum level.

$$J_0 = \frac{em_e k_B T}{2\pi^2 \hbar^3} \int_{E'_n + E''_n/2}^{\infty} D(E, U_a) \ln(1 + \exp((E_F - E)/(k_B T))) dE . \quad (22)$$

These electrons are scattered at the cathode and determine the losses. Then the efficiency can be taken in the form $\eta = J_g / (J_g + J_0)$. For an accurate determination, the total current density should be used. For level l below level *n-k*, the balance conditions take the form

$$\begin{gathered} P_{0l} N_l + P_{1l}(N_l - N_1) + P_{2l}(N_l - N_2) + .... + P_{(l-1)l}(N_l - N_{l-1}) + \\ + P_{(l+1)l}(N_l - N_{l+1}) + ... P_{nl}(N_l - N_n) = 0 \end{gathered} . \quad (23)$$

They are composed of the probabilities of transition per second from level $l$ to all other levels. For higher levels, it is necessary to add the probability of transition from the cathode to the level. Thus, for the upper and the next level.

$$\begin{gathered} P_{0n} N_n + P_{1n}(N_n - N_1) + P_{2n}(N_n - N_2) + \\ | + .... + P_{(n-1)n}(N_n - N_{n-1}) = J_{gn} S / e \end{gathered} , \quad (24)$$

$$\begin{gathered} P_{0(n-1)} N_{n-1} + P_{1(n-1)}(N_{n-1} - N_1) + .... + \\ + P_{(n-2)(n-1)}(N_{n-1} - N_{n-2}) + P_{n(n-1)}(N_{n-1} - N_n) = J_{g(n-1)} S / e \end{gathered} . \quad (25)$$

Equations (23)–(25) are $n$ equations for determining n occupancies. In them, it is assumed that $P_{kl}=P_{lk}$. If the probabilities are known, the occupancies can be determined. For the probabilities, we have $P_{kl} = P_{kl}^0 (1 + f_{BE}(\omega_{kl}, T))$. Thermalization is important if the order $\hbar\omega_{kl}$ is on the order of or less than $k_B T$, i.e., at room temperature, this corresponds to the far-IR and THz ranges. In our case, $\Delta E$ is on the order of 1 eV, and we can consider $P_{kl} = P_{kl}^0$.

To determine the transition probabilities, consider a QW with an electron at level $k$ for $t<0$. In this case, we assume $U_a=0$. The right barrier is infinite, and the level is real. The wave functions $\psi_k(x)$ for this problem are found quite simply. Then, let an anode voltage act for $0<t<t_0$. This results in a finite right barrier. It is necessary to find the solution $\psi_k(x,t,U_a)$ of the non-stationary SE with the conditions $\psi_k(x,0,U_a) = \psi_k(x)$. This problem has been solved in [39]. Then the transition probability over time $t$ is

$$P_{kl}^0(t) = \int \psi_k(x,t,U_a) \psi_l^*(x) dx . \quad (26)$$

Since we are interested in transitions with high energies, we will not consider thermalization further. In the absence of a field (at low temperatures), transitions from low to higher levels are absent. For a single level in a well, the probability of transition to distant wells is small. It is convenient to use the flight times through the barriers between wells to estimate the probabilities (26).

## 8. RADIATION POWER, VAC AND EFFICIENCY OF QQL

In the case of several wide, shifted QWs, there may be many levels in them. Let it be possible for tunneling to occur from the cathode into several of the first QWs. If the FE is large and the barrier height is small, there are many transition scenarios and many oscillation frequencies. Coherent and incoherent oscillations are possible. The levels in the wells depend on the configuration of the neighboring wells. The simplest models are obtained by neglecting this dependence. The simplest calculation is obtained with single-level QWs, when one QW corresponds to one level of the figure. 5. The most likely transition from it is to a neighboring QW. To do this, the electron must tunnel through the barrier. If the level lifetime is long, transitions to the second, third, etc. QW on the right are possible. It is necessary to relate the tunneling times to the levels, which are determined by their energy and the level lifetimes. The process becomes more complex if there are several levels in each well. Ignoring complex scenarios, let us consider a simple model. Let the bottom of each well be lowered relative to the previous one by $\Delta V = 2$ eV, and the level of the well be $E_1 \approx 0.1$ eV relative to its bottom. The levels of neighboring wells are shifted by $\Delta V = 2$ eV. Let FE = 7 eV. Then electrons from the cathode can tunnel resonantly only into the first three wells. According to (6), we calculate three current densities: $J_1$, $J_2$, $J_3$. The first well contributes an additional electron flow to the second well, and it contributes to the third. The total coherent current density QQL is $\tilde{J} = J_1 + J_2 + J_3$. The radiated power is defined as $P = \tilde{J} U_a S$. On the other hand, the power through the lifetimes of the levels is defined as $P = 2\omega S\left[N_1 E_1'' + N_2 E_2'' + N_3 E_3'' + N(n-3)\hbar/(2\tau)\right]$. We assume that the lifetimes of the levels and their occupancies *N* for all wells are equal $\tau = \hbar/(2E'')$, except for the first three, are the same, and coherent emission of quanta of a single frequency $\omega$ occurs. Now, by calculating the total current *J*, we will find the efficiency $\eta = \tilde{J}/J$. The total current $J = \tilde{J} + \Delta J$ consists of the current in the first three QWs and the current from the regions with non-resonant energies $J = \tilde{J} + \Delta J$. At $\Delta V = 3$ eV, we obtain two levels and two first wells into which current from the cathode is possible. This case corresponds to FL of QQL. Tunneling from the cathode into the first QWs is strong if the first barrier is narrow. In this case, the population of the first level is high. The first barrier can be narrowed by applying a positive potential $U_g$ to the first QW relative to the cathode. In this case, we obtain a triode with controlled generation.

Let each well correspond to two levels. Then a transition from the upper level to the lower one is possible, either with the emission of a quantum or with a non-radiative transition. Transitions from both levels are also possible via tunneling and the emission of quanta into neighboring wells on the right, followed by the emission of quanta. In this case, it is possible to emit quanta of different frequencies. The QQL current–voltage characteristics have the following features. At low voltages, the current is small, and the current–voltage characteristic starts from zero. The initial increase in current is provided by ordinary tunneling. Then, resonant tunneling and radiation appear: coherent and incoherent. At this stage, the current reaches its maximum and exhibits oscillations associated with rearrangements of the radiative levels [41]. Further increase in the current voltage removes the structures from resonant tunneling and leads to a decrease in the current. Specifically, the lowering of the levels in the first QW below the bottom of the CZ cathode prevents resonant transitions in them. At the same time, the intensity of conventional tunneling transitions also decreases. This is due to the fact that at a low depth of the first wellt, the barrier between it and the cathode turns into a narrow triangular bevel from the cathode's FL into the wellt. Although the barrier is extremely narrow and highly transparent, a step appears. The reflection coefficient from such a step approaches minus one as the depth of the pit increases. This effect is similar to the effect of optical reflection from a plate with a large DC. If a very thin, almost transparent layer with negative permittivity is formed on the plate (for example, a thin, translucent metal layer emitting a barrier), and behind it there is a thick layer of dielectric with high permittivity, the reflection coefficient $R=(1-\varepsilon)/(1+\varepsilon)$ will be close to –1. Such a current–voltage characteristic with a falling section can be used for generation or amplification when QQL is connected to a transmission line. In this case, such amplification or generation will also be accompanied by the emission of photons of different frequencies due to the rearrangement of levels when the anode voltage changes. Due to the high anode voltage, QQL can exhibit a very large negative differential resistance, which is advantageous for amplification and generation purposes in resonators and transmission lines.

## 9. CONCLUSION

The paper proposes and investigates new QQL structures based on MIM technologies. Their advantage lies in the fact that QWs can be shifted both to levels on the order of eV, which corresponds to generation in the optical and UV ranges, and to fractions of an eV, which corresponds to the IR and THz ranges. Models for calculating metastable levels are proposed. Strictly speaking, the levels must be calculated for the entire structure; however, with wide barriers, they can be calculated for individual QWs. A model is proposed for determining the population of levels with and without taking into account thermalization, since the temperatures are assumed to be low and the quanta to be large. Accounting for the final temperatures requires the use of the general formula for thermofield emission and diffusion-ballistic transport in the structure. The total density of the tunnel current $J$ in all cases is determined by calculating the transparency $D(E)$ and integrating over energy. The quantity $P=JU_a$ is the total power density, which includes the generation power and the loss power. Since tunneling processes occur without a change in energy, this power is either completely dissipated at the anode (if no levels arise in the structure) or partially dissipated and partially emitted in the form of radiative transitions between levels. The splitting of levels in a heterostructure also occurs due to the Rashba spin-orbit interaction effect in an asymmetric internal field. The broadening of levels allows transitions to occur with a small level detuning. At high temperatures, dissipative tunneling with a reduction in efficiency is possible in long QQLs. The emergence of QQL radiation is associated with coherent emission of transitions between equidistantly spaced levels. MIM structures are promising for QQL in the short-wavelength range.

**Funding**

The work was carried out with the financial support of the Ministry of Education and Science of the Russian Federation within the framework of the state assignment (FSRR-2026-0006).

**Conflict of interest**

The author of this work declares that he has no conflicts of interest.

REFERENSES

## Figures

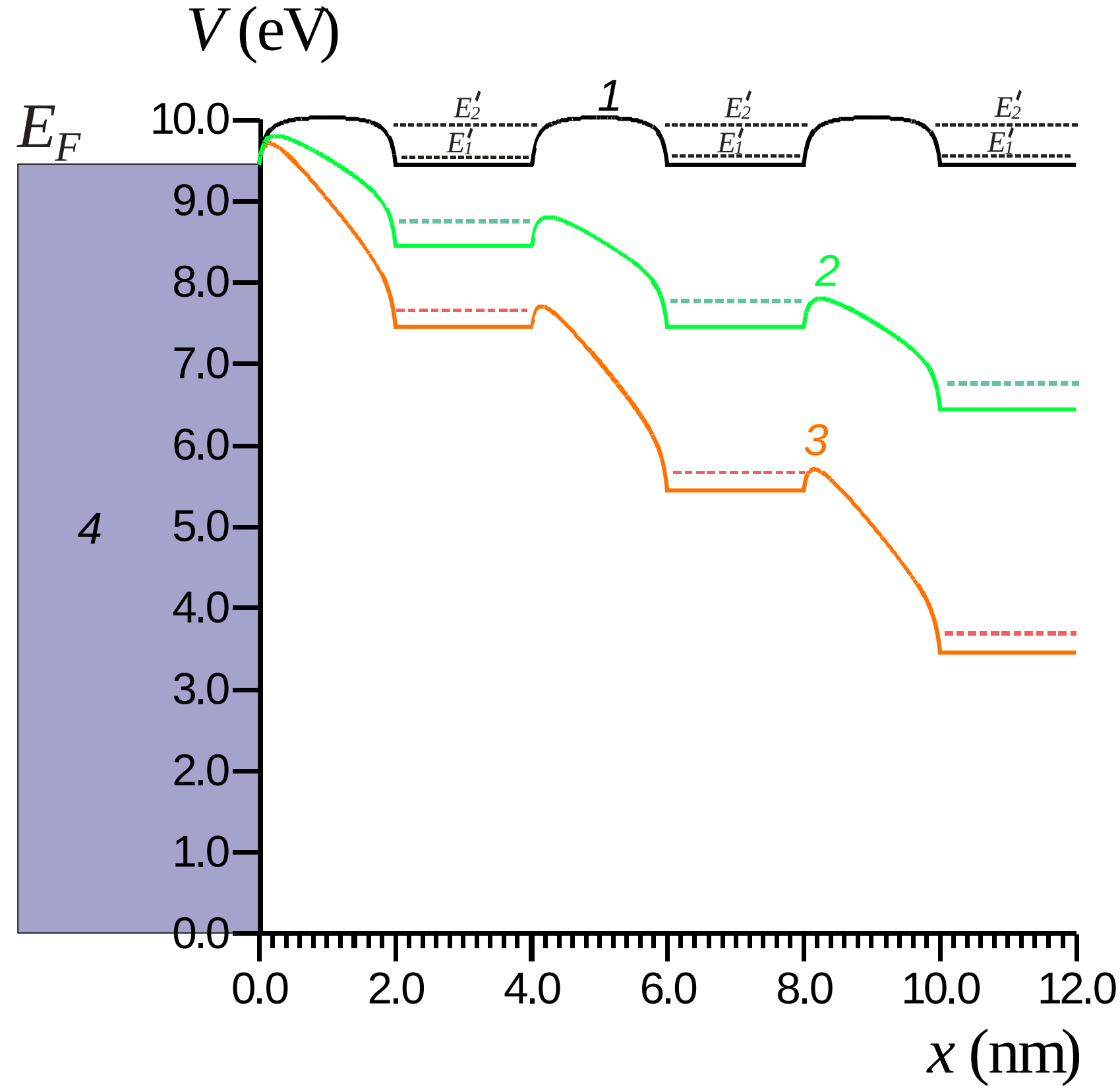

Fig. 1. The profile of the quantum potential $V(x)$ at several wells $d_w$=2 nm and barrier sizes $d_b$=2 nm for PB W=4.2 eV, FE $E_F$=9.45 eV without anode voltage (curve 1), with a 1 V anode voltage drop at each barrier (2) and 2 V on each barrier (3). 4 – CZ of cathode

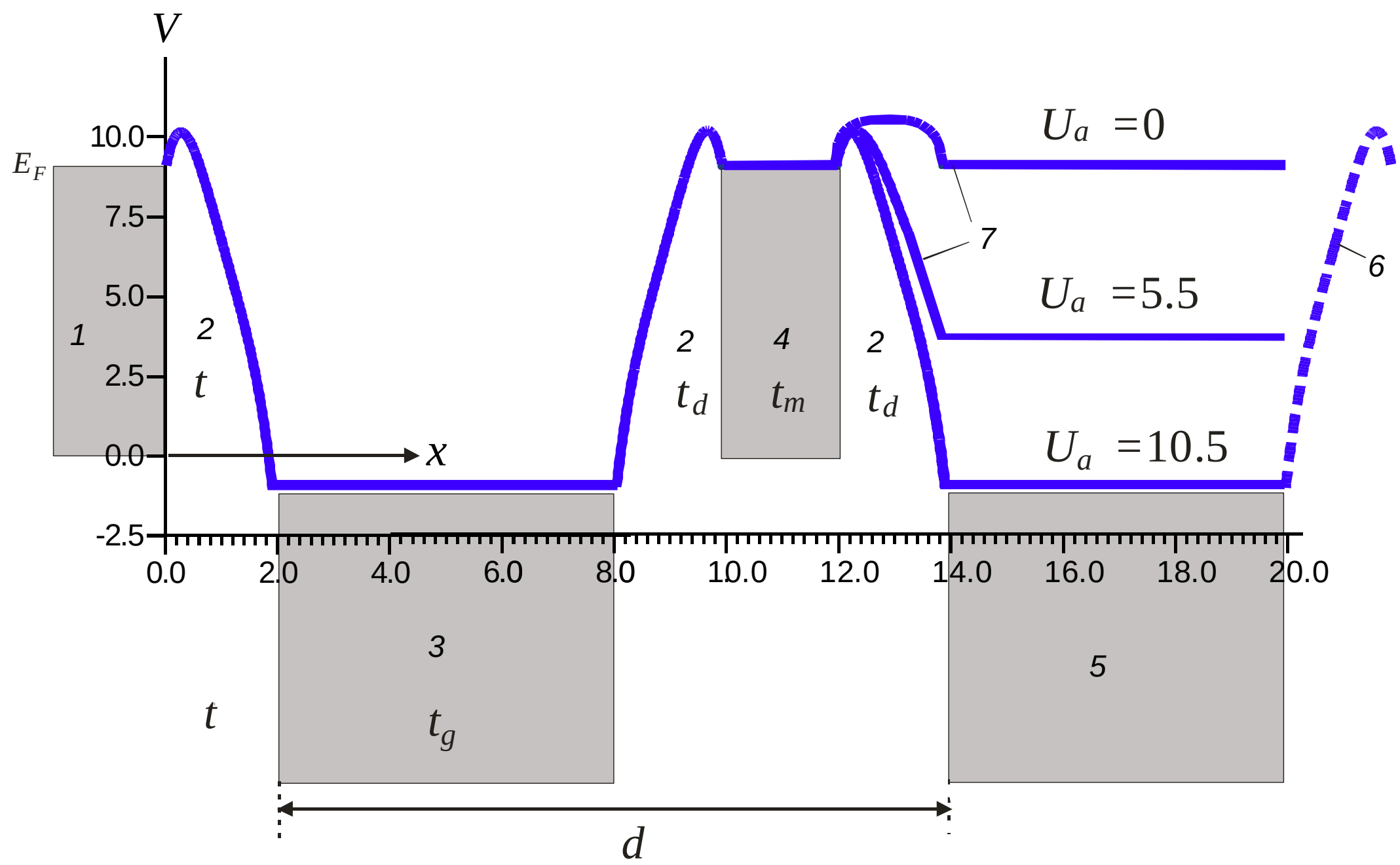

Fig. 2. Distribution of the quantum potential $V(x)$ (eV) in a Pb-diamant heterostructure at a fixed voltage across the wells, depending on the length (nm) for one period. The gray areas indicate the CZ cathode (1), the grids (3), the electrode in the barrier (4), and the anode (5); 2 indicates the barrier regions with dielectric layers. Voltages $U_a$=0, 5.5, and 10 (V) are considered. $E_F$=9.45 eV. The periodic continuation is shown as 6.7 – the potential at the anode in the case of one period.

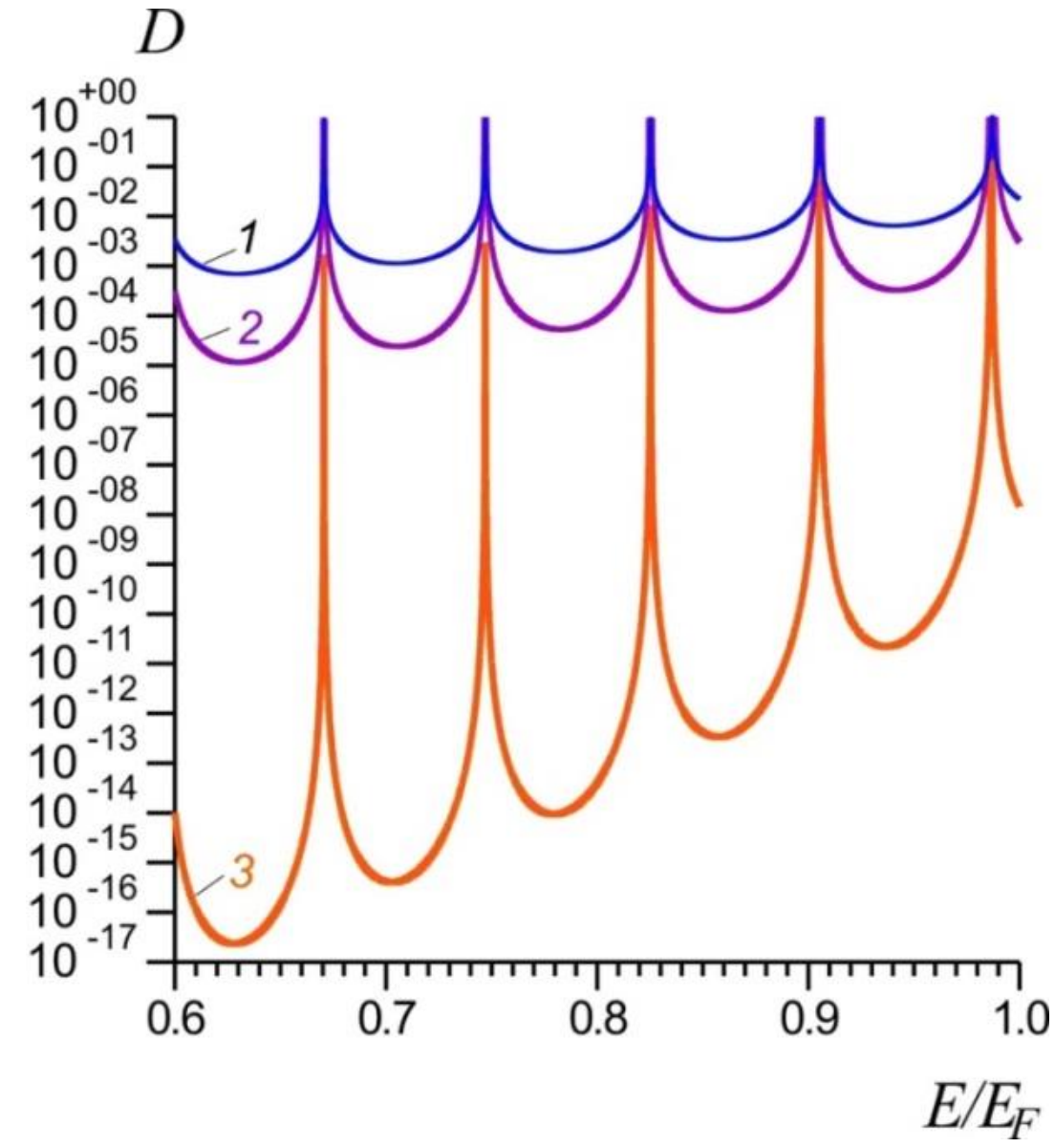

Fig. 3. Tunneling coefficient $D$ in one (curve 1), two (2), and four (3) QWs with a width of 6 nm at $t_b$ = 0.5 nm, $U_a$ = $U_g$ = 10 V.

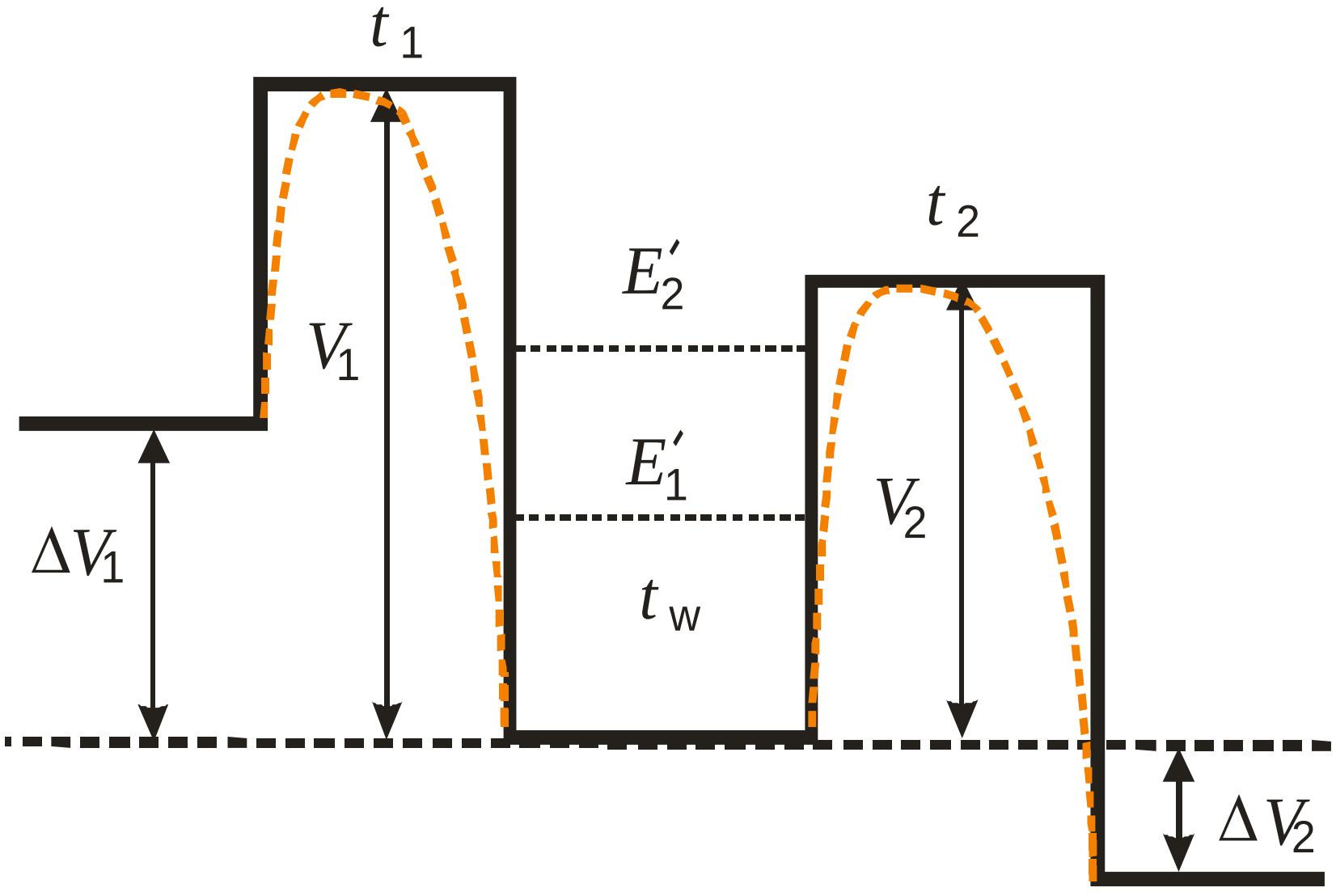

Fig. 4. The structure of the QW potential between a well shifted by a potential $\Delta V_1$, a barrier $V_1$, and a barrier $V_2$, followed by a well shifted by a potential $-\Delta V_2$

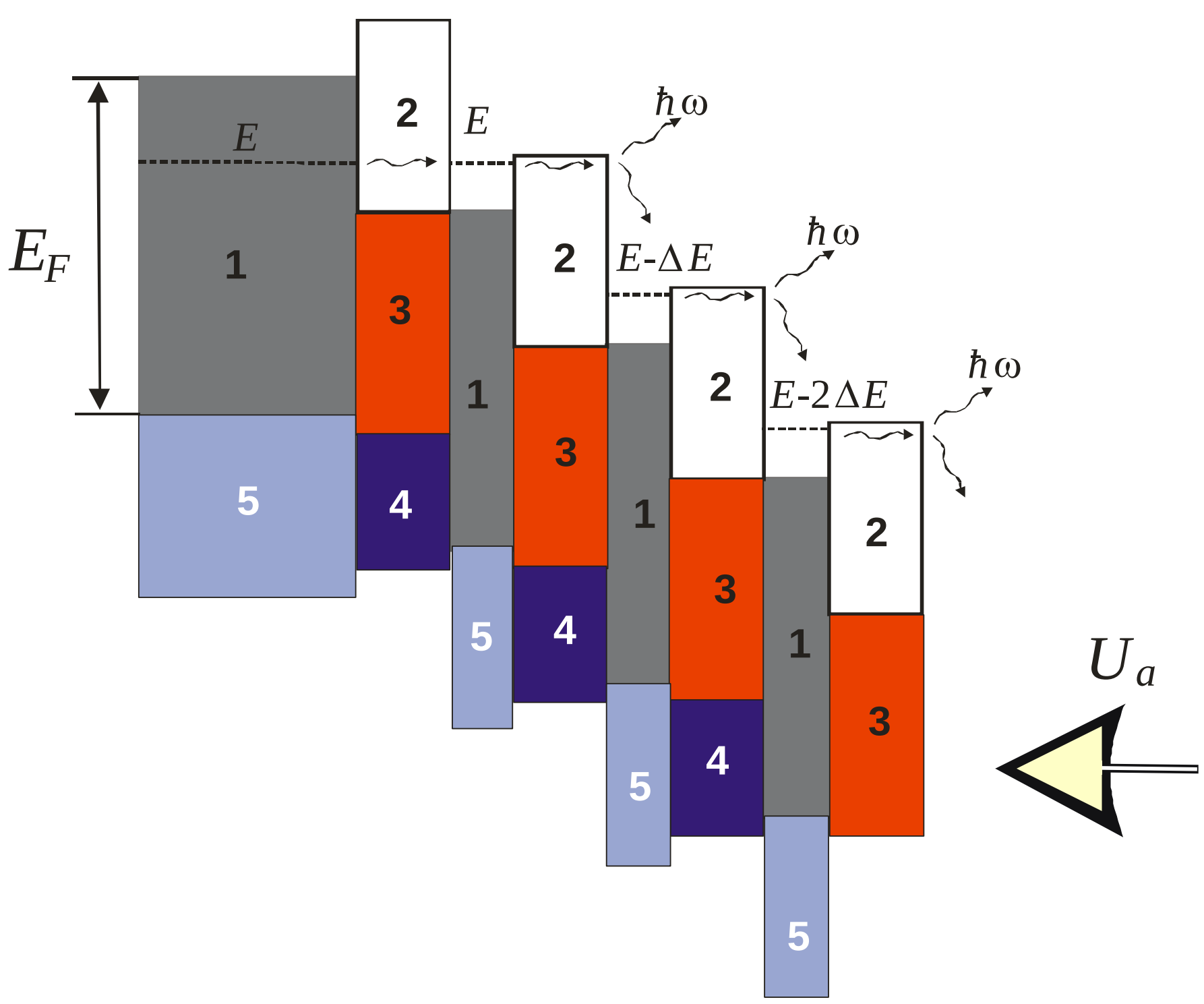

Fig. 5. QQL scheme with a single level in narrow QWs. 1 – CZ of the cathode and QW, 2 – empty CZs of the dielectric barriers, 3 – forbidden bands of the dielectric barriers, 4 – valence bands of the dielectric barriers, 5 – valence bands of the cathode and QW. All wells are shifted by an energy $\Delta E$